\documentclass[%
 reprint,
 amsmath,amssymb,
 aps,
]{revtex4-1}

\usepackage[a4paper, margin=2cm]{geometry}

\usepackage{physics}
\usepackage{amsmath,amssymb}
\usepackage{mathtools}
\usepackage{graphicx}
\usepackage{dcolumn}
\usepackage{bm}
\usepackage{color,soul}
\usepackage{array}
\usepackage{amsthm}
\usepackage{subcaption}
\usepackage{algorithm}
\usepackage{algpseudocode}
\usepackage{wrapfig}
\usepackage{qcircuit}
\usepackage{xcolor}
\usepackage{chngcntr}
\usepackage{setspace}
\usepackage[colorlinks,linkcolor=blue,anchorcolor=red,citecolor=red]{hyperref}
\usepackage[capitalize]{cleveref}
\usepackage{siunitx}

\begin{document}
	
	\title{Quantum optimisation via maximally amplified states}
	
	\author{T. Bennett}
 	\email{tavis.bennett@research.uwa.edu.au}
	\affiliation{Department of Physics, The University of Western Australia, Perth, Australia}

	\author{J. B. Wang}
	\email{jingbo.wang@uwa.edu.au}
	\affiliation{Department of Physics, The University of Western Australia, Perth, Australia}

	\date{\today}
	
	\begin{abstract}
	
	This paper presents the Maximum Amplification Optimisation Algorithm (MAOA), a novel quantum algorithm designed for combinatorial optimisation in the restricted circuit depth context of near-term quantum computing. The MAOA first produces a quantum state in which the optimal solutions to a problem are amplified to the maximum extent possible subject to a given restricted circuit depth. Subsequent repeated preparation and measurement of this maximally amplified state produces solutions of the highest quality as efficiently as possible. The MAOA performs considerably better than other near-term quantum algorithms, such as the Quantum Approximate Optimisation Algorithm (QAOA), as it amplifies optimal solutions significantly more and does so without the computationally demanding variational procedure required by these other algorithms. Additionally, a restricted circuit depth modification of the existing Grover adaptive search is introduced. This modified algorithm is referred to as the restricted Grover adaptive search (RGAS), and provides a useful comparison to the MAOA. The MAOA and RGAS are simulated on a practical vehicle routing problem, a computationally demanding portfolio optimisation problem, and an arbitrarily large problem with normally distributed solution qualities. In all cases, the MAOA and RGAS are shown to provide substantial speedup over classical random sampling in finding optimal solutions, while the MAOA consistently outperforms the RGAS. The speedup provided by the MAOA is quantified by demonstrating numerical convergence to a theoretically derived upper bound.
	
	\end{abstract}
	
	\maketitle
	
	\section{Introduction}
    
    Quantum computing provides a paradigm that exploits quantum-mechanical principles, such as superposition and entanglement, to solve computational problems far more efficiently than current or future classical computers~\cite{Shor1994, Grover1997, Nielsen2010, Montanaro2016, Childs2018}. A significant potential application of quantum computing is in finding high quality solutions to combinatorial optimisation problems. This class of problems appears often and across a broad range of contexts, for example, in commercial settings such as vehicle routing, in financial settings such as portfolio optimisation, and even in medical research such as protein folding. Often these problems have solution spaces that grow exponentially with increasing problem size, which makes finding the optimal solutions for large problems classically intractable \cite{intractability}. Quantum computers have the ability to operate on these exponentially large solution spaces in quantum parallel by using superposition of computational basis states, one assigned to each solution, the total number of which grows exponentially in the number of qubits. This ability is what allows optimisation algorithms such as the Grover adaptive search (GAS) \cite{GAS2,GAS,GAS3} to deliver significant speed up in finding optimal solutions within unstructured solution spaces relative to a classical random-search or exhaustive search procedure. 
    
    However, due to environmental noise, decoherance, and an insufficient number of qubits for error protection, current and near-term quantum computers cannot produce effective and accurate computation at large circuit depths \cite{Preskill2018, NISQ_limitations}. Since the aforementioned GAS algorithm necessarily requires large circuit depths of $\mathcal{O}(\sqrt{N})$, where $N$ is the size of the solution space, it is therefore not likely to be implementable on near-term quantum devices. Consequently, much of the recent quantum algorithm research and development has been focused on achieving quantum advantages while restricted to small circuit depths. One such restricted circuit depth algorithm, focused specifically on combinatorial optimisation, is the Quantum Approximate Optimisation Algorithm (QAOA) \cite{QAOA, farhi2015quantum, farhi2019, Morales2020, Zhou2020}. 
    
    The initial motivation behind the alternating operator ansatz which underpins the QAOA is related to the quantum adiabatic theorem \cite{Adiabatic_theorem} and its use in the quantum adiabatic algorithm (QAA) \cite{Adiabatic_QC}. The quantum adiabatic theorem states that a quantum system will remain in its ground state if its Hamiltonian changes sufficiently slowly with time. The QAA involves preparing a system in the ground state of a known Hamiltonian, then slowly evolving it to the ground state of a Hamiltonian that encodes the cost function of the optimisation problem. The QAOA seeks to approximate this process by instead applying these two Hamiltonians on a quantum circuit in alternating fashion, where the application times are controlled and tuned via a classical optimisation process. Alternating application of the Hamiltonians is in essence an amplitude amplification process~\cite{Hadfield2019}. The classical optimisation process seeks to improve the expectation value of solution quality as measured from the final amplitude amplified state, hence increasing the probability that a measurement of this state produces a high quality solution.
    
    The Quantum Walk Optimisation Algorithm (QWOA) \cite{Marsh2019,marsh2020combinatorial} was developed as a generalisation of this process, where it was recognised that application of the two Hamiltonians was essentially equivalent to a continuous time quantum walk over a connected graph, and a quality-dependent phase shift applied to each solution state on the graph. This theoretical framework has proven extremely useful in subsequent research and indeed in the research presented in this paper. For example, the QWOA has been shown to provide a significant improvement in performance over the QAOA for a portfolio optimisation problem when restricting the quantum walk mixing process to just a subset of valid solutions \cite{portfolio}.
    
    Despite its recent popularity, we identify three primary weaknesses of the QAOA, each of which reduces its efficiency in finding optimal solutions to large combinatorial optimisation problems. These issues are explored in depth later in this paper, though can be summarised as follows. The first limitation is that optimising for expectation value of quality does not provide maximum speed-up in finding optimal solutions. A much more effective approach would focus directly on maximising amplification of optimal solutions. The second limitation is that the QAOA does not take advantage of the degrees of freedom available within the QWOA framework. Instead, it provides no treatment to the quality distribution prior to application of quality-dependent phase shifts, and in its typical form, it makes use of the transverse field operator. By generalising to the QWOA framework, and exploring its degrees of freedom, we see that optimal solutions can be amplified significantly more by first transforming the quality distribution to one which is binary (combined with a Grover mixer). The third and perhaps most significant limitation of the QAOA is that its variational procedure is very computationally expensive, as with any high-dimensional optimisation process \cite{LONG2019108}, and that this computational expense prevents the method from providing practical speedup. By exploring and addressing each of these three weaknesses we arrive at the underlying mechanics of a novel algorithm, introduced in this paper as the Maximum Amplification Optimisation Algorithm (MAOA).
    
    The remainder of this paper is organised as follows. In \cref{sec:Justifying_MAOA} the QWOA framework is reviewed and the aforementioned limitations of the QAOA are explored within this framework. This exploration provides a natural introduction to the amplification process which is central to the MAOA. In \cref{sec:Connection_with_Grovers}, the connection between this amplification process and Grover's search is outlined. This relationship is central in developing the MAOA. In \cref{sec:MAOA}, the MAOA is introduced and presented in detail, followed by a discussion on the GAS and RGAS in \cref{sec:GAS}, as these will form a relevant baseline for the comparison of algorithm performance. Numerical simulations, results and analysis are then presented in \cref{sec:Simulation} and \cref{sec:MAOA_analysis}. 
    
    \section{Justifying the Maximum Amplification Optimisation Algorithm}
    \label{sec:Justifying_MAOA}
    
    \subsection{The QWOA framework}
    
    A detailed description on the theoretical framework of the Quantum Walk Optimisation Algorithm (QWOA) was given in a previous paper \cite{CVRP}; it is included here for clarity and completeness. Formally, we consider a mapping $f: \mathbb{S} \longrightarrow \mathbb{R}$, which returns a measure of the quality associated with each possible solution in the solution space $\mathbb{S}$, where $\mathbb{S}$ has cardinality $N$. 
    
    The starting point of the QWOA is a quantum system with $N$ basis states, one for each solution in $\mathbb{S}$, initialised in an equal superposition,
    \begin{equation}
        \ket{s} = \frac{1}{\sqrt{N}}\sum_{x \in \mathbb{S}}\ket{x}.
        \label{eq:eqsuperpos}
    \end{equation}
    
    \noindent This initial state is then evolved through repeated and alternating application of the \emph{quality-dependent phase-shift} and \emph{quantum-walk-mixing} unitaries. The quality-dependent phase-shift unitary is defined as
    \begin{equation}
        U_Q(\gamma_j) = \exp(-\text{i} \gamma_j Q ),
    \end{equation}
    \noindent where $\gamma_j \in \mathbb{R}$ and $Q$ is a diagonal operator such that $Q\ket{x} = f(x) \ket{x}$. The quantum walk mixing unitary is defined as
    \begin{equation}
        U_W(t_j) = \exp(-\text{i} t_j A ),
    \end{equation}
    
   \noindent where $t_j > 0$, and $A$ is the adjacency matrix of a circulant graph that connects the feasible solutions to the problem, i.e. the graph contains $N$ vertices, one for each solution in $\mathbb{S}$ and the vertices/solutions are connected according to the adjacency matrix, $A$. Note that the Laplacian matrix of the graph could also be used, but it would produce equivalent behaviour. The graphs are selected to have circulant connectivity, because all circulant graphs are diagonalised by the Fourier transform and hence can be efficiently implemented on a quantum computer~\cite{Qiang2016, Loke2017b, Zhou2017, Qiang2021}.
    
    The first unitary $U_Q$ applies a phase-shift at each vertex proportional to the quality of the solution at that vertex, with the proportionality constant given by the parameter, $\gamma_j$. The second unitary $U_W$ can be understood as performing a quantum walk over the graph for time $t_j$, mixing the amplitudes across vertices. Following the mixing of phase-shifted amplitudes across the vertices of the graph, constructive and destructive interference will result in quality-dependent amplitude amplification, controlled by the parameters $\gamma_j$ and $t_j$. Application of $U_Q$ and $U_W$ is repeated $r$ times, resulting in a final state of the system given by
    \begin{equation} \label{eq:QWOA}
        \ket{\bm{\gamma}, \bm{t}} = U_W(t_r) U_Q(\gamma_r)...U_W(t_{1})U_Q(\gamma_{1}) \ket{s},
    \end{equation}
    
    \noindent where $\bm{t} = (t_1, t_2, ..., t_r)$ and $\bm{\gamma} = (\gamma_1, \gamma_2, ..., \gamma_r)$. 
    
    By tuning the parameters $\bm{\gamma}$ and $\bm{t}$, it is possible to amplify the amplitudes corresponding with high quality solutions, and therefore increase the probability of a measurement of the system collapsing it into a high quality solution. The process of tuning the parameters, also known as the variational procedure, is conducted iteratively through the use of a classical optimisation algorithm which takes as its objective function the expectation value of the $Q$ operator:
    \begin{equation}
    \label{eq:expectation}
        c(\bm{\gamma}, \bm{t}) = \bra{\bm{\gamma}, \bm{t}}Q\ket{\bm{\gamma}, \bm{t}}.
    \end{equation}
    
    The QWOA framework also assumes there exists an indexing algorithm which provides a one to one mapping from indices $\in \{0,1,...,N-1\}$ to solutions in the solution space. This allows for the quantum walk and graph connectivity to be restricted to just the space of valid solutions. Note also that the QAOA is contained within the QWOA framework, where application of the transverse field/mixing operator of QAOA is equivalent to a quantum walk over a hypercube \cite{Marsh2019}.
    
    \subsection{The issue with optimising for expectation value of quality}
    \label{sec:expectation_value} 
\begin{figure}[H]
        \centering
        \includegraphics[width=0.95\columnwidth]{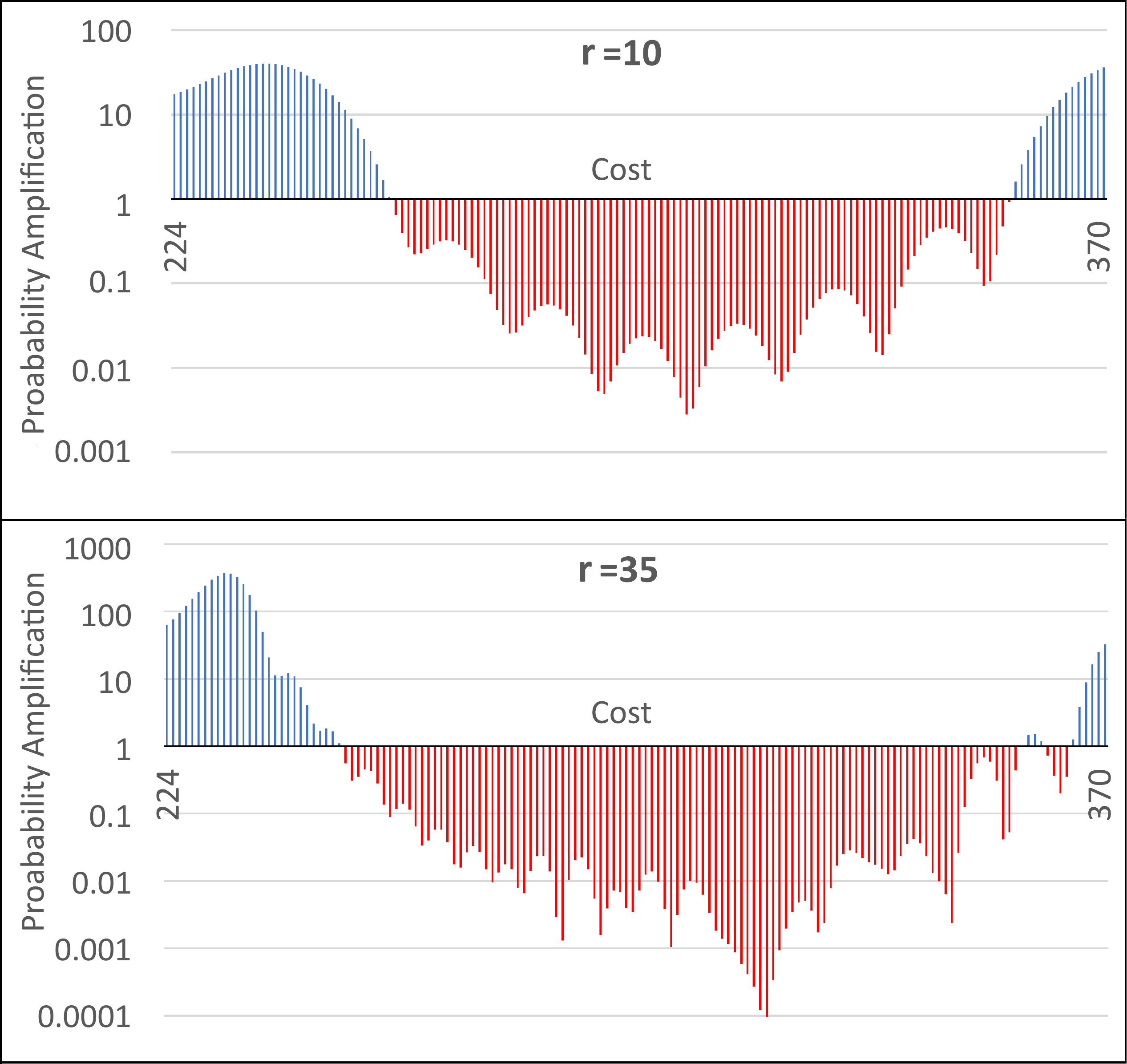}
        \caption{Probability amplification of solutions to a vehicle routing problem as a function of their cost after 10 and 35 QWOA iterations \cite{CVRP}. Note that the optimal (lowest-cost) solutions are amplified considerably less than the more numerous sub-optimal solutions. In addition the least-optimal (highest-cost) solutions are amplified in spite of their low quality.}
        \label{fig:CVRP_amplifications}
    \end{figure}

    Many combinatorial optimisation problems exhibit solution qualities with normal-like distributions (examples of which are shown in \cref{sec:Simulation}), in which case sub-optimal solutions will significantly outnumber the optimal solution(s). Since the variational procedure within the QWOA framework operates on the expectation value for quality, $c(\bm{\gamma}, \bm{t})$, it tends to amplify a large group of sub-optimal solutions more than a less numerous group of optimal solutions. The reasoning for this is subtle. The expectation value for quality would clearly be optimised if the optimal solutions are completely amplified. However, with restricted circuit depths, the maximum possible amplification is limited. In this context of limited and quality-dependent amplification, the expectation value of quality is optimised when favouring amplification of the sub-optimal solutions, simply due to their superior number within the solution space. Moreover, it is also possible for the least-optimal solutions to be amplified as a secondary effect, since these are less numerous and do not exert enough influence on the expectation value of quality to favour their suppression during the optimisation process. Both of these effects were seen clearly in our previous QWOA simulation results \cite{CVRP}, included here in \cref{fig:CVRP_amplifications}.
    
    This clearly demonstrates that the QWOA/QAOA variational procedure does not, in general, maximise amplification of the optimal solutions. On the other hand, the variational procedure produces an amplified state in which the expected value of quality has been optimised. This may not seem like an issue, because when measuring from such an amplified state, we expect to find a solution of reasonably high quality. This is not, however, the most effective approach when searching for the optimal solution(s) to a large problem. Instead, the focus should be on producing a quantum state in which the optimal solutions are maximally amplified.
    
    \subsection{Amplification of optimal solutions as a more effective metric}
    
    The context of the QWOA and QAOA is in solving large and hence classically-intractable combinatorial problems with near-term and hence restricted circuit depth quantum computers. For large problems and restricted circuit depths, it is not possible for the optimised expectation value of quality to converge to the highest solution quality, instead, it will converge to a reasonably high, sub-optimal quality. This is because the amount of amplification available is limited by the restricted circuit depth, and for the expectation value of quality to converge to that of the optimal solutions, these optimal solutions require extremely large amounts of amplification, from what is initially only a very small component of the initial equal superposition. If the goal of these near-term algorithms is to find a solution of reasonably high quality, then the variational procedure will be successful in producing a quantum state which, when measured, will regularly produce reasonably high quality solutions. However, this is not, and should not be the goal. Firstly, the variational procedure is computationally expensive, and secondly, producing a reasonably high quality solution is trivially easy through classical means. For example, the process of randomly sampling 1,000 solutions and taking from these the best quality solution, is expected to produce a solution with quality in or near the top 0.1\% of all solutions. Instead, the goal of QWAO and QAOA should be to find an optimal solution, or at least a solution which is very close to being optimal (near-optimal).
    
    Amplification produced in the QWOA amplified states is necessarily limited by restricted circuit depth. As such, a QWOA amplified state is likely to require repeated preparation and measurement in order to produce an optimal or near-optimal solution, since with limited amplification, measurement of these solutions remains relatively unlikely. The amplification of a solution represents the rate at which it will be measured from the amplified state relative to the rate it will be measured via classical random sampling. So the amplification of near-optimal and optimal solutions is the single metric most relevant to the speed-up provided by measuring any particular amplified state. It is for this reason that the following section is focused on maximising amplification of optimal solutions.
    
    \subsection{Effect of graph structure and quality degeneracy on amplification of a single optimal solution}
    
    In order to understand how the two primary degrees of freedom within the QWOA framework, the graph structure used for the quantum walk and the quality distribution, affect the amplification of the optimal solution in a solution space, a meticulous numerical investigation has been carried out, discussed in detail in \cref{sec:AppendixA} and \cref{sec:AppendixB}, with the important results summarised as follows:
    
    \begin{itemize}
        \item Increased degeneracy in a quality distribution increases the amplification of the single optimal solution. As such, a binary marking function applied to a quality distribution produces the largest amplification of a single marked solution for any graph. Note that degeneracy here refers to degeneracy in the non-optimal qualities. A binary marking function is defined here as a function which transforms the quality distribution into one of marked (quality $=1$) and unmarked (quality $=0$) solutions. In general the binary marking function will mark all solutions with quality superior to some specified threshold value, but in this case the single optimal solution is the only marked solution.
        \item A binary marking function also produces maximum amplification of the single marked solution with repeated applications of the same phase-shift and walk-time parameters, reducing the optimisation landscape for the classically tuned parameters to one that is 2-dimensional for any number of iterations.
        \item Specifically in the case of a binary marking function, of all the investigated graphs, the complete graph produces the maximum amplification of the single marked solution. Note that the complete graph is the graph in which each vertex/solution is connected to every other.
    \end{itemize}
    
    Besides the fact that a binary-marked complete graph produces the highest amplification of the optimal solution with repeated application of the same parameter pairs, it also has another significant advantage, discussed in the next section.
    
    \subsection{The reduced complete graph}
    
\begin{figure}[ht]
        \centering
        \includegraphics[width=0.95\columnwidth]{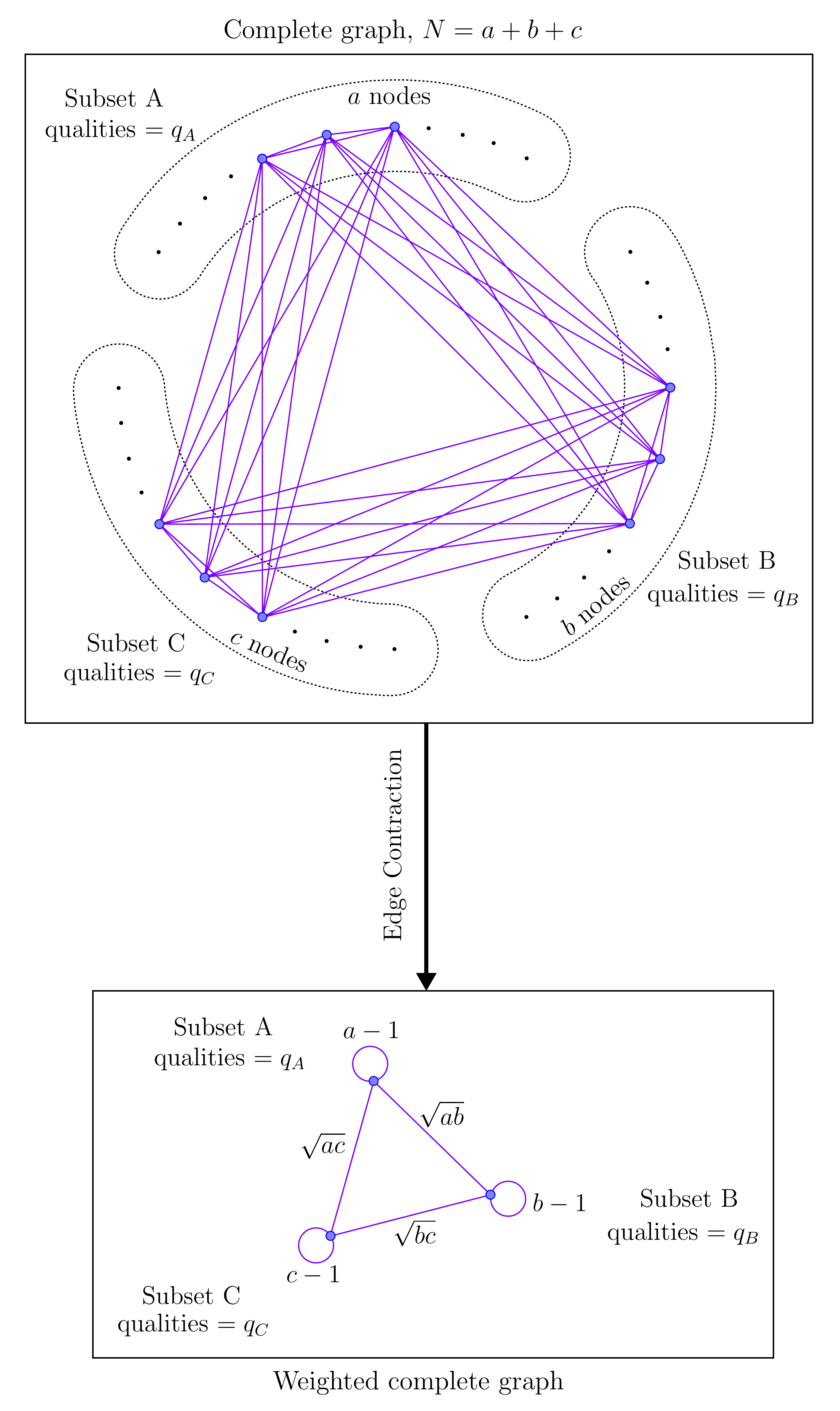}
        \caption{An illustration of the edge contraction process for a quality distribution containing 3 distinct qualities, distributed over a complete graph. This shows how the behaviour of an arbitrarily large complete graph can be greatly simplified.}
        \label{fig:Edge_contraction}
    \end{figure}

    The complete graph presents a unique opportunity in that its behaviour can be greatly simplified when there is degeneracy in the distribution of qualities across its vertices~\cite{Marsh2021, Grover_Equivalence}. The degenerate vertices can be combined through an edge contraction process to produce a graph with significantly fewer vertices which produces equivalent behaviour with respect to amplitude amplification within degenerate groups. This simplified graph will be referred to from here on as the reduced graph. The reason degenerate vertices can be combined in this way is because each of them are functionally equivalent within the graph. They each receive identical phase shifts and have identical connectivity with regards to neighbouring vertices of each quality.  \cref{fig:Edge_contraction} illustrates the edge contraction process and shows how it produces a reduced graph. Note the illustration is specifically for an example solution space containing solutions with 3 distinct qualities. The solutions space can therefore be divided into 3 subsets, $S_A$, $S_B$ and $S_C$, with respective qualities, $q_A$, $q_B$ and $q_C$ containing respectively, $a$, $b$ and $c$ solutions each. In any case, the reduced graph is a weighted graph where each vertex represents one group of degenerate vertices from its parent graph. The weight of each regular edge corresponds with the square root of the number of edges connecting the respective degenerate groups in the parent graph. The weight of each self loop is equal to one less than the number of vertices in the respective degenerate group. The self loops are necessary because they account for the mixing of amplitude that occurs within vertices of the same group.
    
    The reduced graph in \cref{fig:Edge_contraction} can be characterised by the following quality operator $Q_3$, adjacency matrix $A_3$, and initial equal superposition state $\ket{s_3}$, where
    
    	\[ Q_3 = \left[ \begin{array}{ccc} q_A & 0 & 0 \\ 0 & q_B & 0 \\ 0 & 0 & q_C \\ \end{array} \right], \;\; A_3 = \left[ \begin{array}{ccc} a-1 & \sqrt{ab} & \sqrt{ac} \\ \sqrt{ab} & b-1 & \sqrt{bc} \\ \sqrt{ac} & \sqrt{bc} & c-1 \\ \end{array} \right],\]
    	
    	\[\ket{s_3} =  \left[ \begin{array}{c} \sqrt{\frac{a}{N}} \\ \sqrt{\frac{b}{N}} \\ \sqrt{\frac{c}{N}} \\ \end{array} \right].\]
    
    This process displays quite clearly a rather significant advantage of the complete graph. Its behaviour with regards to amplitude amplification within degenerate groups of vertices is insensitive to solution placement within the solution space or across the vertices of the graph. This hints at the possibility for analytically derived optimal parameters for amplification into an optimal set of solutions, or at the very least, parameters which are also insensitive to solution placement/ordering with respect to their resulting amplitude amplification.
    
    \subsection{Optimal parameters for a binary marking function on a complete graph}
    
    Up to this point, the focus has been on just a single optimal solution and its amplification. However, in reality, there is no way to know the number of solutions marked by a binary marking function on an unknown solution space, or similarly the number of solutions in the most-optimal partition for some other partitioning of the solution space. As such, the focus will now be on investigating amplification of some unknown fraction of marked solutions.
    
    The two partition or binary marked problem on a complete graph is characterised by the following reduced graph adjacency matrix, quality operator and initial state, where $m$ represents the number of marked vertices (solutions), and $N$ represents the total number of vertices (solutions), namely
        \[ Q_2 = \left[ \begin{array}{cc} 1 & 0 \\ 0 & 0 \\ \end{array} \right], \;\; A_2 = \left[ \begin{array}{cc} m-1 & \sqrt{m(N-m)} \\ \sqrt{m(N-m)} & N-m-1 \\ \end{array} \right] ,\]
        \[\ket{s_2} =  \left[ \begin{array}{c} \sqrt{\frac{m}{N}} \\ \sqrt{\frac{N-m}{N}} \\ \end{array} \right].\]

    \noindent Consider a single iteration amplified state, 
        \[\ket{s_2'} = U_W(t) U_Q(\gamma) \ket{s_2}.\]
    \noindent Substituting $m=\rho N$, where $\rho$ is the ratio of marked solutions (which is  also the initial marked solution probability), it is possible to derive an expression for the probability contained in the marked solutions of the amplified state. Dividing by $\rho$, then taking the limit for small $\rho$, the final expression for amplification of the marked solutions after a single iteration becomes
    \begin{equation}
    \label{eq:amplification_r=1}
        3+2(\cos{Nt}(\cos{\gamma}-1)-\cos{\gamma})-2\sin{Nt} \sin{\gamma}
    \end{equation}
    \noindent which takes a maximum value of 9 for $\gamma=\pi$ and $t=\frac{\pi}{N}$.
    
    As described previously, and demonstrated within \cref{sec:AppendixB}, amplification of a single marked vertex on the complete graph is maximised by repeated application of the same parameters. Since the above expression for amplification is independent of $m$, and any binary marked complete graph can be characterised by $Q_2$, $A_2$ and $\ket{s_2}$, the amplification of marked vertices should be independent of whether we are looking at a single marked vertex, or any arbitrary number of marked vertices, so long as the ratio of marked vertices is small. So amplification should also be maximised by repeated application of the same parameters for complete graphs with multiple marked vertices. As such, repeated application of the derived parameters, $\gamma=\pi$ and $t=\frac{\pi}{N}$, is expected to maximally amplify marked vertices on the binary marked complete graph, for arbitrary $r$. 
    
\begin{figure}[ht]
        \centering
        \includegraphics[width=0.95\columnwidth]{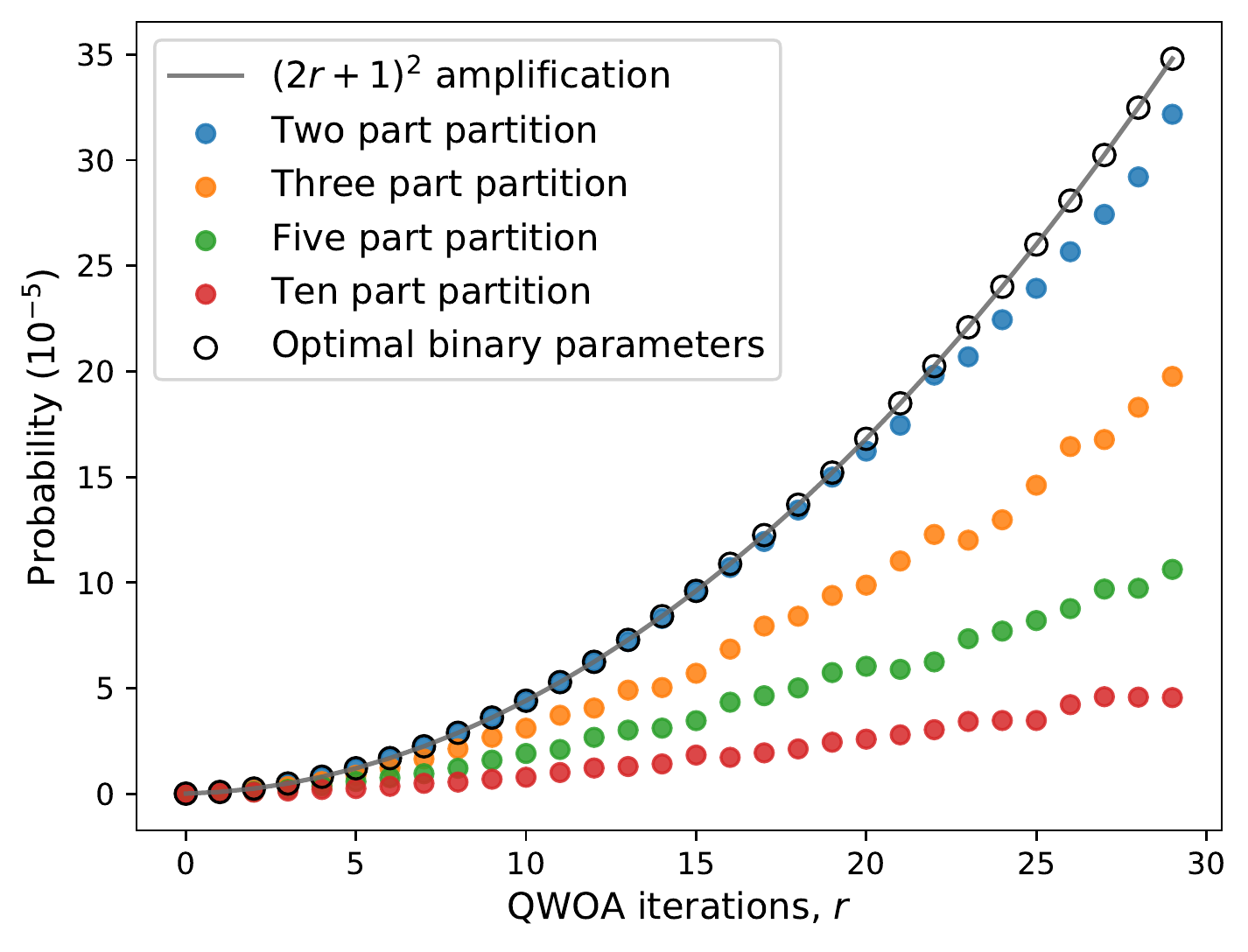}
        \caption{Optimised amplification of 10 marked vertices/solutions on a complete graph of $10^8$ total vertices/solutions, with increasing iterations of the QWOA process. We see that a small number of optimal solutions can be amplified significantly more by assigning them into the marked set of a binary partition, than for any other partitioning of the solution space with less degeneracy in the non-optimal solutions.} \label{fig:various_partition_optimisation}
    \end{figure}

    In order to show that the binary marking function remains the most effective at amplifying a small group of marked vertices on a very large complete graph and across a range of iteration numbers $r$, the performance of various partitions of the complete graph will be assessed. Namely, 2, 3, 5 and 10 part partitions will be assessed. In addition, the derived parameters, $\gamma=\pi$ and $t=\frac{\pi}{N}$, will be applied to the binary partitioned complete graph to ensure that they do indeed produce optimal amplification. In each case, the graph will have a total of $N=10^8$ vertices, with 10 vertices in the marked partition (with quality 1). The remaining vertices will be partitioned into equal sized groups to achieve the required total number of partitions and each group assigned with a single quality from those distributed uniformly over the interval [0,1]. Note that amplified probabilities for the marked vertices are computed using \cref{eq:QWOA} with the adjacency matrix, quality operator and initial state taken as those for each respective reduced graph. The process for optimising amplification into the marked vertices for a given number of iterations and a given partitioned graph consists of randomly generating 10,000 sets of $2r$ initial parameters. The three sets of parameters that produce maximum initial amplification are taken as the initial values in a Nelder-Mead optimisation procedure. This process was repeated 24 times and the maximum probability from the 72 optimised results was taken as the final most-optimal probability. The results of this analysis are shown in \cref{fig:various_partition_optimisation}, where it is clear that the binary (two-part) partition performs the best, and that the rate of amplification of the optimal solutions decreases significantly as we tend towards the typical QWOA process with increasing numbers of partitions (i.e. decreasing degeneracy). The solid curve shows the maximal amplification given by $(2r+1)^2$, which fits the observed maximum amplification, a fact that will be addressed in \cref{sec:Connection_with_Grovers}. Repeated application of the derived optimal parameters also clearly matches with the results from the optimisation procedure, at least up until $r=15$, after which point the optimisation procedure is outperformed by the derived parameters, likely because the optimisation procedure is not rigorous enough to find the global maxima in the higher dimensional optimisation landscapes.
    
    It is important to note that, for a given circuit depth, characterised by $r$ QWOA iterations, maximum amplification of a small group of high quality solutions can be achieved by using a binary marking function over a complete graph and repeatedly applying the derived set of parameters, $\gamma=\pi$ and $t=\frac{\pi}{N}$. This is the primary mechanism underlying the MAOA, and allows the variational procedure of the typical QWOA or QAOA approach to be avoided entirely. To understand how this process can be implemented within a generalised optimisation procedure, it is first important to understand how it is related to Grover's search.
    
    \section{Connection with Grover's search}
    \label{sec:Connection_with_Grovers}
    
    \subsection{Grover's rotation and diffusion operators}
    
    To understand how the MAOA works, it's useful to first understand how the alternating application of continuous-time quantum walks over a complete graph and quality dependent phase shifts with a binary marking function are related to Grover's search \cite{OG_grovers}. Note that the following is outlined in further detail by \citet{Grover_Equivalence}.
    
    Consider a binary marking function which takes as its inputs a threshold quality, $T$, and a solution quality, $f(x)$, and returns a 1 if the solution quality is superior to the threshold quality or a 0 otherwise. Application of this marking function would transform the diagonal $Q$ operator to one with eigenvalues of 1 for marked states and 0 for non-marked states. When combined with the optimal phase shift parameter, $\gamma=\pi$, this transforms the quality-dependent phase-shift unitary, $U_Q(\gamma)$, into an operator which applies a $\pi$ phase shift to all marked states. Note that this is functionally equivalent to the Grover rotation operator, which applies a $\pi$ phase rotation to the marked state(s) \cite{OG_grovers}. 
    
    The adjacency matrix of the complete graph can be expressed as, $A=N\ket{s}\bra{s}-\mathbb{I}$, so the quantum walk unitary applied for time $t=\frac{\pi}{N}$ can be expressed as
    \begin{equation}
        U_W(\frac{\pi}{N}) = \mathrm{e}^{-\text{i} \frac{\pi}{N} A} = \mathrm{e}^{-\text{i} \frac{\pi}{N} (N\ket{s}\bra{s}-\mathbb{I})} = \mathrm{e}^{\text{i} \frac{\pi}{N}} (\mathbb{I}-2\ket{s}\bra{s}).
    \end{equation}
    This is equivalent to the Grover diffusion operator \cite{OG_grovers} up to a global phase, and hence produces equivalent behaviour with regards to mixing of states.
    
    Given that the two QWOA unitaries applied in sequence and with the derived parameters, $\gamma=\pi$ and $t=\frac{\pi}{N}$, are equivalent to a single iteration of Grover's search, and both processes operate on the initial equal superposition, then $r$ such QWOA iterations produces the same amplitude amplification of the marked solutions as a Grover's search terminated after $r$ rotations. The key difference between the two processes is that Grover's search aims for complete convergence and requires large circuit depths, where as, with restricted circuit depths, the MAOA will terminate the process early and produce only partial convergence into the marked solutions. 
    
    Even though it has been demonstrated that this process of maximum amplification is functionally equivalent to a truncated Grover's search, it is also important to note that Grover's search, truncated or not, is not on its own a generalised optimisation procedure. The Grover's search procedure will make use of the amplification process to accelerate the search for a predefined element or group of elements from a larger space, but for a combinatorial optimisation problem, the defining features or qualities of an optimal solution or group of solutions is not known. It will therefore require additional work to incorporate this maximum amplification process into a generalised optimisation procedure, the framework for which will be explored in detail in the following sections.
    
    \subsection{The low-convergence regime of Grover's search}
    
\begin{figure}[ht]
        \centering
        \includegraphics[width=0.95\columnwidth]{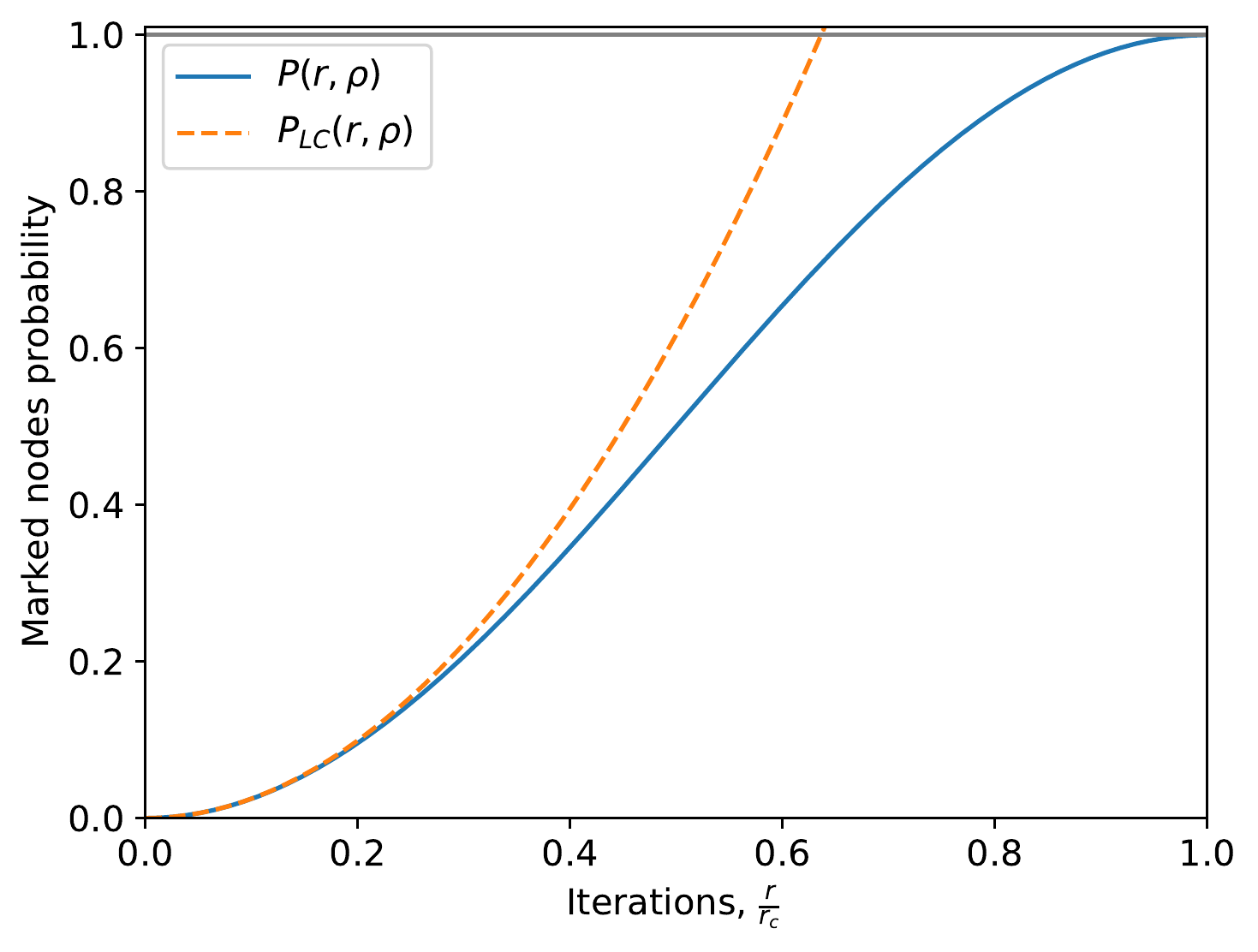}
        \caption{The amplified probability of a small group of marked solutions with increasing number of Grover rotations $r$ relative to $r_c$.}
        \label{fig:grovers_curve}
    \end{figure}

    The amplified probability of the marked states during a Grover's search depends only on the number of completed rotations ($r$) and the ratio of the marked solutions to the total solution space ($\rho=\frac{m}{N}$) \cite{Grover_probability_equation, GAS}, as given by
    \begin{equation}
        P(r,\rho) = \sin((2r+1)\arcsin(\sqrt{\rho}))^2.
        \label{eq:Grovers_prob}
    \end{equation}
    In the limit of small $r$ and $\rho$, \cref{eq:Grovers_prob} reduces to
    \begin{equation}
        P_{LC}(r,\rho) = \rho (2r+1)^2,
        \label{eq:low_convergence_prob}
    \end{equation}
    which gives the probability of measuring marked solutions when the convergence into these states is low. This low-convergence estimate is accurate to within $1\%$ when the amplified probability is less than $\frac{1}{40}$. In \cref{fig:grovers_curve}, we plot these probability expressions with increasing rotation counts, $r$, relative to the rotation count required for complete convergence, i.e. $$r_c=\frac{\pi}{4\arcsin{\sqrt{\rho}}}-\frac{1}{2}.$$

    The region within which the low-convergence approximation is accurate shall be referred to from here on as the low-convergence regime. When applying a certain rotation count, $r$, a sufficiently small marked-vertex ratio will guarantee that the amplified state will be one that exists within this regime, and hence one that is maximally amplified, since for larger marked-vertex ratios, the marked nodes probability will fall below that predicted by \cref{eq:low_convergence_prob}. The amplification applied to the marked solutions within the low-convergence regime can be read directly from \cref{eq:low_convergence_prob}, where $\rho$ is the initial marked-solution probability and hence $(2r+1)^2$ is the amplification relative to the marked-solution probability in the non-amplified state. Once in this regime, adjusting the threshold so as to further reduce the marked ratio will not result in any further amplification of the marked states, but will instead only reduce the total size of the marked set. Note that at this point it should be apparent why the $(2r+1)^2$ amplification curve was included in \cref{fig:various_partition_optimisation}, as it shows perfect agreement with the amplification produced by the application of the derived optimal parameters to the binary-marked complete graph.
    
    \section{The Maximum Amplification Optimisation Algorithm}
    \label{sec:MAOA}
    
    In summary of what has so far been established, for a given circuit depth characterised by $r$ QWOA iterations, a binary marking function on the complete graph produces the maximum possible amplification of the marked states via repeated applications of the QWOA parameters, $\gamma=\pi$ and $t=\frac{\pi}{N}$, so long as the selected quality threshold for the marking function produces a marked ratio small enough to guarantee that the amplified state is in the low-convergence regime of a functionally equivalent Grover's search terminated after $r$ rotations. With respect to finding the optimal solution(s), repeated preparation and measurement of this maximally amplified state represents the maximum speedup available at the specified restricted circuit depth. Much of this section will therefore be focused on presenting a method to reliably and efficiently find a quality threshold which will produce this maximally amplified state.

    Note that whether operating within the QWOA framework or that of the truncated Grover's search, the amplification process remains functionally equivalent, so from this point on, the Grover framework will be adopted, i.e. $r$ will be referred to as the number of rotations.
    
    \subsection{Threshold response curves}
    
\begin{figure}[ht]
        \centering
        \includegraphics[width=0.95\columnwidth]{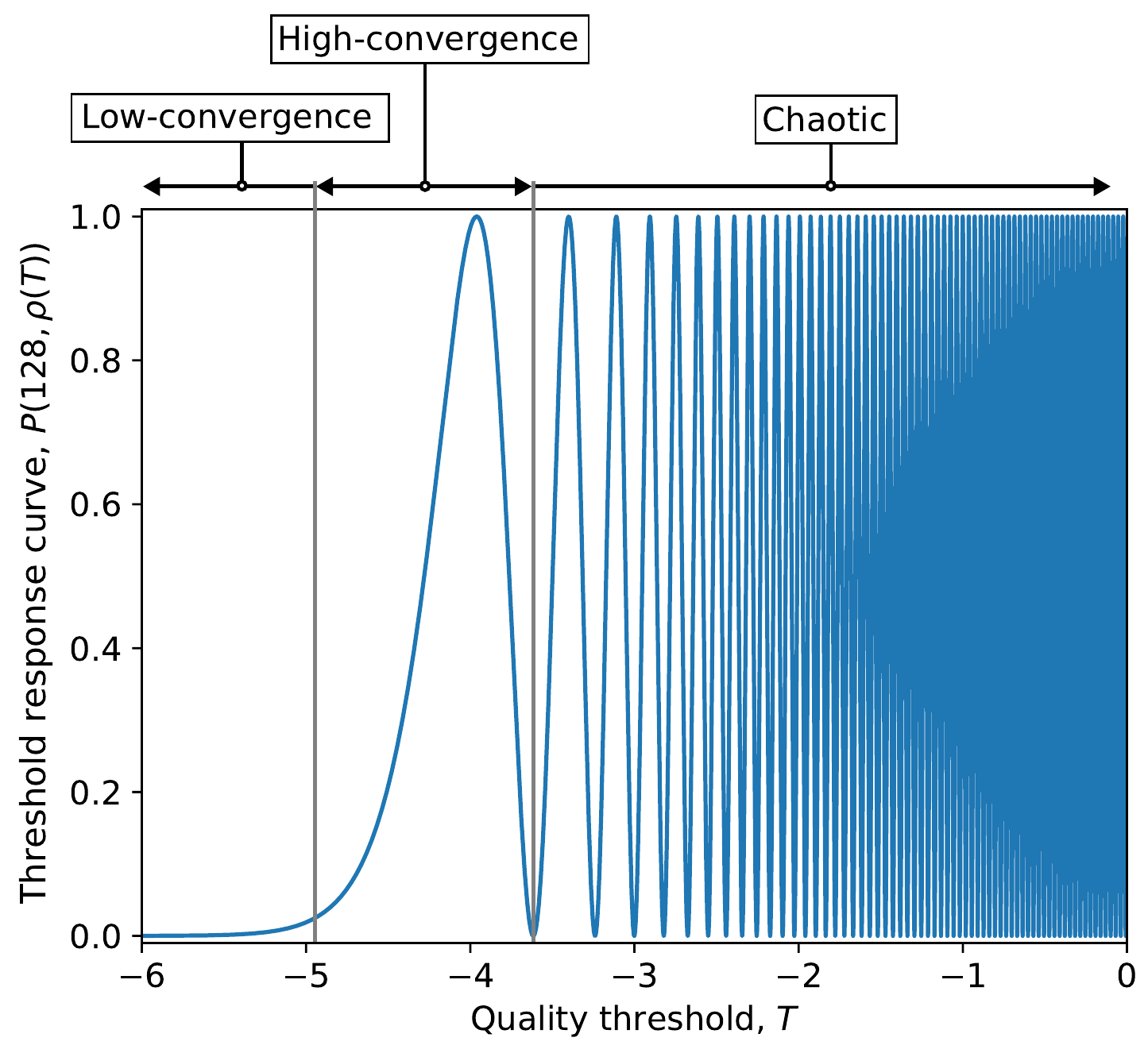}
        \caption{Threshold response curve for 128 rotation ($r=128$) amplified states over a solution space with qualities distributed as per the standard normal distribution. The low-convergence, high-convergence and chaotic regimes are all labelled accordingly.}
        \label{fig:r128_response}
    \end{figure}

    In order to understand how the MAOA locates a maximally amplifying quality threshold, it is first useful to introduce the concept of a threshold response curve, which quantifies, for a fixed number of rotations, $r$, how the probability of measuring a marked solution varies with the quality threshold, $T$. The threshold response curve is given by \cref{eq:Grovers_prob} as $P(r,\rho(T))$ where $\rho(T)$ is the marked-solution ratio produced by the marking function. The threshold response of a system represents the rate of successful measurements of marked solutions. A strong response would occur where the probability of measurement is close to $1$. An example threshold response curve for $r=128$ is shown in \cref{fig:r128_response}. This curve is the response of a system which has a quality distribution matching that of the standard normal distribution, i.e. mean quality of 0, standard deviation equal to 1, and with a sufficiently large number of solutions such that the distribution of solution qualities is approximately continuous. This response curve is for a minimisation problem, i.e. the marked set is all solutions with qualities less than $T$. 
    
    In general, the total combined number of peaks and troughs at either side of the median is equal to the rotation count, $r$, so the response curve in \cref{fig:r128_response} contains $64$ peaks and $64$ troughs. It is useful to define three different regions or regimes within the threshold response curve, each of which is labelled in \cref{fig:r128_response}. The chaotic regime refers to the range of thresholds within which there is a tightly spaced fluctuation between high and low response. In terms of a traditional Grover's search, this is where the number of rotations significantly exceeds the minimum number required for complete convergence, $r>2r_c$. The high-convergence regime refers to the region around the most-optimal quality at which we see peak response. Again, in terms of a traditional Grover's search, this is where the number of rotations first approaches that required for complete convergence, $0.1r_c<r<2r_c$. Finally, the low-convergence regime refers to the region of qualities in which solutions are too few in number for $r$ rotations to be sufficient to produce high-convergence. Note that this low-convergence regime is identical to that defined earlier, where maximum amplification of marked solutions is achieved, and corresponds with $P(r,\rho(T))<\tfrac{1}{40}$ and $r<0.1r_c$.
    
    So the goal of the first part of the MAOA is to find a threshold, $T$, which is located within the low-convergence regime of the threshold response curve $P(r,\rho(T))$, where $r$ corresponds with the restricted circuit depth at which the final process of repeated state preparation/measurement is to be carried out. Lowering the threshold and monitoring the response of the system at the final value of $r$ is not likely to be a practical method, as it requires navigation of the chaotic regime in a controlled fashion, which contains $\frac{r}{2}-1$ peaks and requires a highly accurate estimate of the median. On the other hand, there exists a much more efficient method to navigate from an initial threshold located loosely around the median on the $r=1$ response curve, through to a final threshold within the low-convergence regime of the final $r$ response curve, doubling the rotation count as required. 
    
    \subsection{Navigating the threshold response curves}
    
\begin{figure}[ht]
        \centering
        \includegraphics[width=0.95\columnwidth]{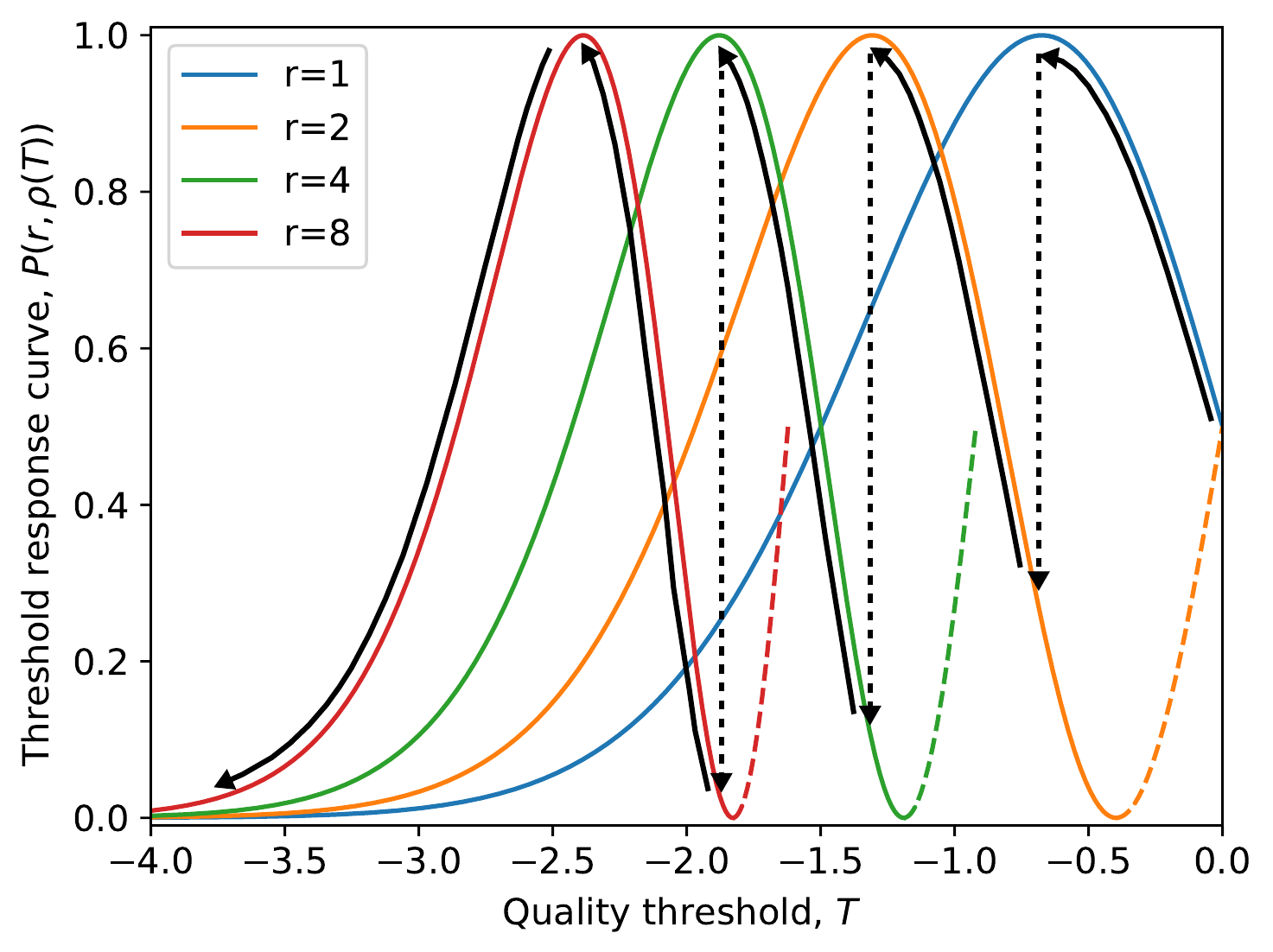}
        \caption{An illustration of how the MAOA navigates across threshold response curves from a threshold near the median for the $r=1$ curve to a final threshold in the low-convergence regime of the $r=8$ curve.}
        \label{fig:r1_to_r8}
    \end{figure}

    The process of navigating from somewhere near the median on the $r=1$ response curve, to the low-convergence regime on the final $r$ response curve is illustrated in \cref{fig:r1_to_r8}, where the final value for $r$ is taken at 8. This iterative process of doubling the rotation count, and moving from the high-convergence peak of one curve to the trough below it on the next curve, allows for the threshold to remain in the well-behaved high-convergence regime. This allows for reliable navigation from one peak to the next until the rotation count corresponding with the desired final circuit depth is reached. For this method, the number of peaks that must be navigated to arrive at the final peak at $r$ grows with $\log_2(r)$. For contrast, navigating the entire threshold response curve at the final $r$ would involve a number of peaks which would grow linearly with $r$, not to mention the fact that any initial threshold would not have a well-defined position on the threshold response curve due to the tight spacing of peaks around the median. The other benefit of navigating through successive response curves is that state preparations/measurements made during the early stages of the process, at low $r$, incur significantly lower computational effort.
    
    It is important to note the necessary connection between doubling the rotation count and the peak to trough relationship between successive threshold response curves, which is not just a coincidental relationship specific to the standard normal distribution. The peak in the high-convergence regime for a given $r$ occurs when the argument of the sine function in \cref{eq:Grovers_prob} is equal to $\frac{\pi}{2}$, conversely, the trough immediately preceding the high-convergence regime occurs when this argument equals $\pi$.  Since this argument is proportional to $r$, the doubling of $r$ at the necessary threshold, transforms from the peak in the high-convergence regime at $r$ to the trough immediately preceding the high-convergence regime at $2r$. This relationship becomes much tighter for large $r$, but still holds sufficiently true at low $r$, as is made evident in \cref{fig:r1_to_r8}.
    
    The peak finding method is presented in detail in \cref{sec:pseudocode} but essentially reduces to recording the number of successful marked solution measurements made in a row at steadily improving thresholds (skipping to the next threshold anytime a measurement is unsuccessful at producing a marked solution). If 20 successful measurements are made in a row, the threshold is held, and the rotation count doubled (this happens most of the time and happens near the peak). If 20 consecutive successful measurements are not made, the process of checking successively improving thresholds is continued until the threshold is past the peak, at which point a weighted mean of the success counts is performed to locate the peak. Assessing whether the peak has been passed, is done initially by making use of the interquartile range of a small sample of the non-amplified solution space, and eventually the peak to peak gaps of the previously navigated threshold response curves. These previous peak to peak spacings are also used to derive an adaptive threshold step-size which is updated throughout, which helps to account for the narrowing of peak to peak gaps with increasing $r$, or any other changes in the distribution at larger separations from the mean.
    
    When arriving at the final peak (within the high-convergence regime of the final $r$) an adaptive search \cite{PAS,HAS} with fixed circuit depth can be performed, knowing that any threshold where $P(r,\rho(T))\leq\frac{1}{40}$ is guaranteed to be located in the low-convergence regime and hence will be one that generates maximum amplification in all marked solutions for the given $r$. This adaptive search procedure is shown in \cref{fig:r1_to_r8} as the final descent on the $r=8$ curve.
    
    An important feature of this threshold finding process is that it operates independently of the exact shape of the underlying distribution in solution qualities. This is important, because if the distribution is known exactly, and can be accurately fit with a relatively efficient sampling process, then a threshold within the low-convergence regime can be trivially deduced. This may be appropriate in some cases, but the reality is, for larger rotation counts, the relevant quality thresholds are located at large deviations from the mean, and the behaviour of the quality distributions in these regions becomes less certain without significant sampling effort. Even for problems with solution spaces which are in general normally distributed, the distributions can vary from the perfect normal distribution at large deviations from the mean, for example, in the vehicle routing problem within \cref{sec:Simulation}. In any case, due to imperfections in a problem's actual quality distribution compared to its idealised form, extrapolating from a classical random-sampling can result in selection of a threshold which is too close or too far from the mean. If too close to the mean, a threshold could be located within a trough in the chaotic regime, exhibiting a threshold response which would be otherwise identical to a maximally amplifying threshold, but one that produces significantly less amplification. If too far from the mean, there may not be any marked solutions at all, or at least so few that they may not be measured at all. 
    
    \subsection{General strategy}
    
    \noindent To summarise, the general strategy for the MAOA consists of two parts: \\
    \\
    \underline{Part 1:} For a given problem and restricted rotation count, $r$, find a quality threshold which produces maximum amplification of the marked solutions (through the application of $r$ Grover rotations). Note that this threshold will be one within the low-convergence regime and hence will amplify marked solutions by the maximum factor of $(2r+1)^2$. \\
    \\
    \underline{Part 2:} Using the quality threshold acquired in part 1, repeatedly prepare and measure the maximally amplified state. The repeated measurement of this amplified state will produce random high quality solutions from the marked set at a rate of approximately $\frac{1}{40}$ or slightly less, depending on the final marked-solution ratio. 
    
    \subsection{Computational effort}
    \label{sec:Comp_effort}
    In order to assess the performance of the MAOA, it is important to be able to quantify the computational effort that it expends. Assuming that the computation of solution qualities is a task that requires a significant fraction of the total computational effort, we will use the number of calls to the quality function, $f(x)$, as a measure of computational effort. Note that in both the Grover and QWOA framework, each time the marked solutions are phase shifted, the quality function is effectively called twice, once to mark the relevant solutions, and once for the uncomputation process, to reset relevant ancillary qubits for subsequent iterations. In addition, once a final solution is measured from the amplified state, its quality must still be computed, adding one more call to quality function. As such, the preparation and measurement of an amplified state incurs a computational effort of $2r+1$. In addition, a classical random-sampling of the solution space incurs a computational effort equal to the number of samples, as it requires only a single call to the quality function for each randomly selected solution. Note that to fully quantify any speedup relative to classical random-sampling, the computational effort involved in the quantum walk/mixing process should also be accounted for. This will not be considered as part of this work, however, the number of walks/mixes is equal to the rotation count, so would pose only a linear overhead on top of the computational effort associated with the objective function calls.
    
    \subsection{Note on  the use of expectation value of quality}
    Recently, \citet{golden2021threshold}, propose a modification of the QAOA which makes use of a Grover Mixer combined with a binary marking function. Note that the Grover mixer is effectively a continuous-time quantum walk over the complete graph \cite{Grover_Equivalence}, and so the underlying mechanics of their algorithm is close to that of the MAOA. The primary differences are that the performance of the amplified state is monitored via the expectation value of quality, the final pair of parameters are left free for a tuning process, and otherwise the parameter corresponding with the Grover mixer is taken as $\pi$ rather than $\frac{\pi}{N}$. As an aside, the periodicity of the walk on the complete graph is such that a walk time of $\frac{(2i+1)\pi}{N}$, where $i \in \{0,1,2,\dots\}$, will produce maximum mixing, but a walk time which is some multiple of $\frac{2\pi}{N}$ will complete an integer number of cycles and produce no mixing and hence no amplification. This can be observed in the amplification expression in \cref{eq:amplification_r=1}. As such, the mixing parameter equal to $\pi$ will function well only for a solution space with an odd number of solutions. 
    
\begin{figure}[ht]
        \centering
        \includegraphics[width=0.95\columnwidth]{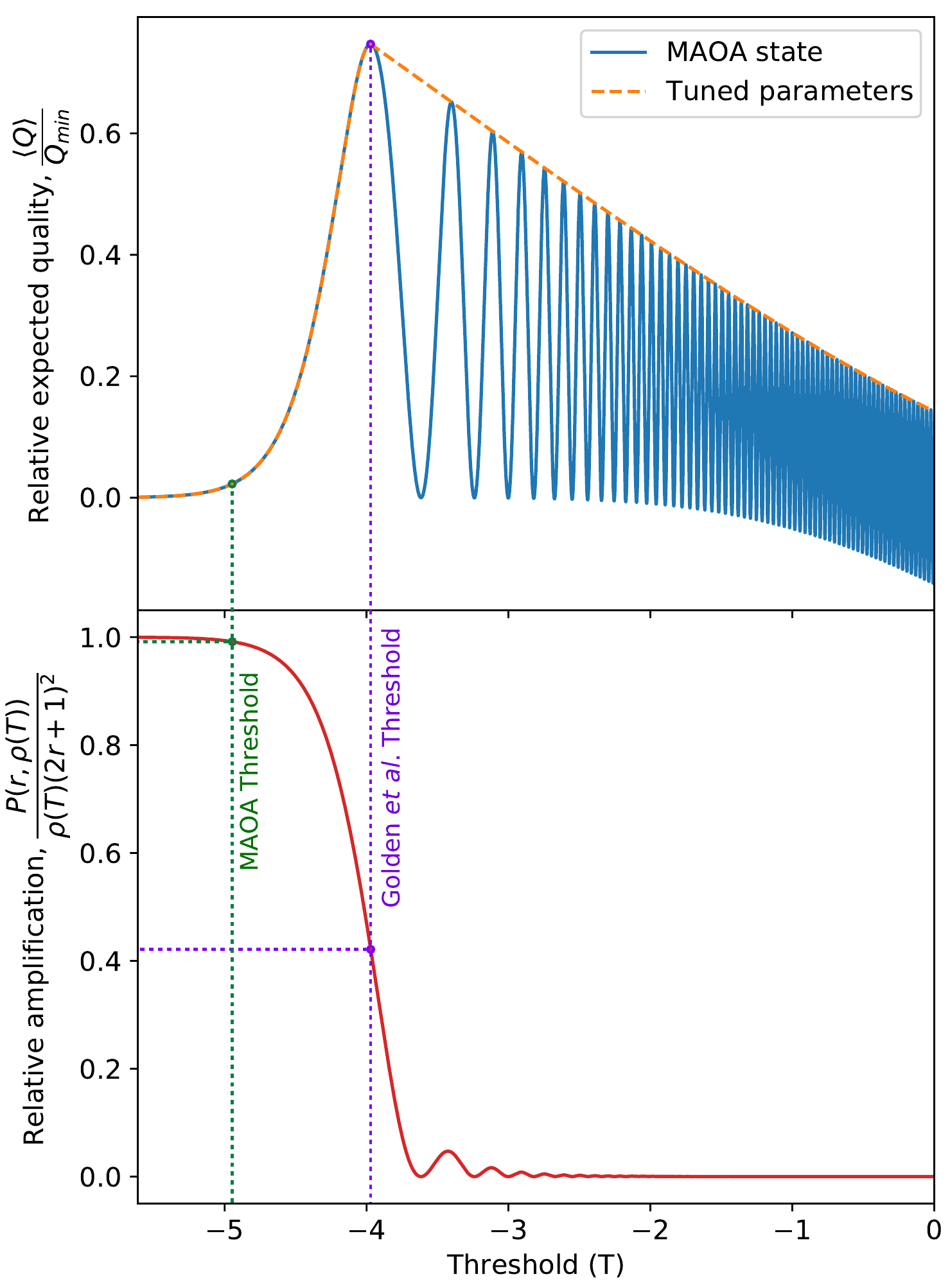}
        \caption{Threshold response for the expectation value of quality over a quality distribution matching the standard normal distribution. Amplification of the marked set is also shown, relative to the maximum possible amplification. This demonstrates that selection of a binary threshold which maximises expected quality produces significantly less amplification of optimal solutions when compared to a threshold selected via the MAOA.}
        \label{fig:expectation_response}
    \end{figure}

    By monitoring the expectation value of quality, \citet{golden2021threshold} choose a threshold for their marking function which maximises the expected quality produced by the amplified state. The issue with this is that, as discussed in \cref{sec:expectation_value}, a threshold which maximises expectation value of quality will not produce maximum amplification of the highest quality solutions. In fact, measurement of a state in which the optimal solution(s) are maximally amplified will only occasionally produce a marked solution (approximately 1 in every 40 measurements or less), and hence will have an expectation value for quality which is close to the mean quality of the solution space. The expectation value is therefore not a useful metric in determining whether a state produces maximum amplification of the optimal solution(s). This is illustrated in \cref{fig:expectation_response}, where a 128-rotation amplified state is prepared over a solution space with qualities distributed as per the standard normal distribution. The expectation value of quality is shown relative to a target quality corresponding to the minimum solution out of 100 million total solutions. The amplification of the marked set is also shown, relative to the maximum possible amplification. 
    
    Note that by tuning the applied parameters, it is possible to produce expected qualities along the upper envelope of the MAOA curve, as shown with the dashed curve in \cref{fig:expectation_response}. This dashed curve matches the observed behaviour shown in Figure 1 of the paper by \citet{golden2021threshold}. With fixed parameters, the 128-rotation amplified state of the MAOA oscillates in accordance with the chaotic regime of Grover's search. As shown in \cref{fig:expectation_response}, the threshold which produces the best expectation value for quality corresponds with an amplification of the marked set which is on the order of only 40\% of the maximum possible amplification. The MAOA threshold, on the other hand, produces essentially the maximum possible amplification. Ignoring the computational effort involved in the tuning of the phase shift and mixing parameters involved in this binary threshold version of QAOA, as well as the process of optimising expected quality, this approach would not be capable of producing the same speedup as the MAOA simply due to the fact that it produces less than half of the maximum possible amplification of the marked solutions.
    
    \section{The Grover adaptive search as a benchmark}
    \label{sec:GAS}
    
    As mentioned in the introduction, a highly effective quantum optimisation algorithm already exists, called the Grover adaptive search (GAS) \cite{GAS2,GAS,GAS3}. However, it is unlikely to be implementable on near-term quantum devices, due to a requirement for large rotation counts of $\mathcal{O}(\sqrt{N})$. Nevertheless, the performance of this algorithm makes for a useful baseline for comparison against the performance of the MAOA. The D{\"u}rr and H{\o}yer (randomised rotation count) variation of the algorithm will be employed for comparison purpose, which is summarised by \citet{GAS}. The GAS effectively uses the Grover's search procedure to amplify the iteratively improving marked set and perform a hesitant adaptive search \cite{HAS}, which is essentially a pure adaptive search \cite{PAS} where the probability of successfully finding an element in the improving set is generally less than 1. 
    
    Since the GAS navigates the solution space in a random manner, there is no way of knowing whether, for a particular rotation count, the chosen threshold falls within the chaotic regime or not, and hence the rotation count must be randomised throughout. As the marked set improves, and hence the marked ratio decreases, the rotation count required to suitably amplify the marked set increases. So although the rotation count needs to be randomised, it is randomly selected from a uniform distribution which has a steadily growing upper-bound. The rate at which this rotation count upper-bound grows is controlled by a parameter, $\lambda$. In their work, \citet{GAS} show that the GAS performs optimally on a solution space with uniformly distributed qualities when $\lambda=1.34$. As it will become clear in \cref{sec:Simulation}, combinatorial optimisation problems often have solution spaces which possess something closer to a normal distribution in qualities. We confirm via simulation that $\lambda=1.34$ remains optimal for normally distributed solution spaces. As such, we will use this value in subsequent simulations. 
    
    
    
    A novel but quite natural modification of the GAS, making it more suitable for the restricted circuit depth context of near-term quantum computation, is simply to place a limit on the maximum allowable rotation count, where the procedure is otherwise identical. This modified version of the GAS will now be referred to as the restricted Grover adaptive search (RGAS). As will be seen in \cref{sec:Simulation}, the RGAS remains effective in providing speedup relative to classical random-sampling in terms of finding optimal solutions. The speedup is related to the limit placed on the rotation count, with larger rotation count restrictions producing better performance, closer to that of the original GAS, and unsurprisingly, smaller restrictions performing worse. The RGAS will therefore make for a useful comparison with the MAOA, as they can both be restricted to the same circuit depth/rotation count, leveling the playing field, and ensuring comparison between two algorithms which are equally well suited to the context of near-term quantum computation. Note that we should expect the MAOA to outperform the RGAS, because the RGAS requires a randomised rotation count, meaning that many of the amplified states will be prepared using smaller rotation counts than the upper limit, and hence won't make use of the maximum possible amplification, whereas, once the final threshold has been located, the MAOA produces the maximally amplified state every time.  
    Note also that the typical QWOA and QAOA procedures have not been included for comparison, since the MAOA achieves significantly more amplification of optimal solutions and does so without the need for a computationally expensive variational procedure to arrive at the optimal phase and walk parameters. The MAOA does require a single dimensional optimisation to arrive at the final quality threshold, but this is readily achievable.
    
    \section{Numerical simulation}
    \label{sec:Simulation}
    
    Now that the Maximum Amplification Optimisation Algorithm (MAOA), Grover adaptive search (GAS) and its restricted circuit depth variant (RGAS) have all been established, their performances will be compared in the context of combinatorial optimisation. Firstly, each of the three algorithms will be applied, via numerical simulations, to a capacitated vehicle routing problem. Next, they will be applied to a portfolio optimisation problem. Lastly, the MAOA and the RGAS will be compared in the limit of arbitrarily large normally distributed solution spaces. In every case, curves for success probability vs. computational effort will be computed from the results of 10,000 simulations. Success probability is defined in each case, but is essentially the probability of having found an optimal solution after a specified amount of computational effort. Each simulation operates over the actual distribution of solution qualities for each particular problem, where each of these distributions is precomputed. The key assumptions underlying each of the simulations are listed below, and are consistent with the dynamics of a truncated Grover's search:
    
    \begin{enumerate}
        \item A given threshold, $T$, produces a marked-solution ratio, $\rho(T)$, which is fully defined by the precomputed quality distribution.
        \item For a given threshold, $T$, and rotation count, $r$, the probability of measuring a marked solution is given by $P(r,\rho(T))$ as per \cref{eq:Grovers_prob}.
        \item When a marked solution is successfully measured, it has an equal probability of being any one of the marked solutions, so the returned solution is randomly selected (uniformly) from the full set of marked solutions for the specified threshold.
        \item The computational effort required for each preparation and measurement of an amplified state is quantified by $2r+1$, as per \cref{sec:Comp_effort}.
    \end{enumerate}
    
    
    The classical method of randomly sampling the solution space in search of optimal solutions, similar to exhaustive-search, is included in all cases as a baseline for comparison, since the performance of random-sampling is well defined and problem independent. We acknowledge, however, that exhaustively searching the solution space for high quality solutions is not the fastest classical approach for most combinatorial optimisation problems. Never-the-less, this comparison allows the speedup of MAOA to be clearly quantified, as discussed in \cref{sec:MAOA_analysis}.
    
    
    \subsection{The capacitated vehicle routing problem}
    
    A detailed theoretical framework for the capacitated vehicle routing problem (CVRP) is presented by \citet{CVRP} and adopted without modification here for the purpose of generating a solution space and corresponding quality distribution for analysis. The problem essentially involves seeking the lowest cost routes for delivering supplies from a central depot to a number of external locations.  The cardinality of the solution space is given by 
    \begin{equation}
        \label{eq:N_CVRP}
        N_{CV}(l) = \sum_{k=1}^{l}{l-1 \choose k-1} \frac{l!}{k!},
    \end{equation}
   where $l$ is the number of locations. For the following simulation, we set $l=10$ giving $N=58,941,091$.
    
    \begin{figure}[ht]
        \centering
        \begin{subfigure}{1\columnwidth}
            \centering
            \includegraphics[width=0.85\columnwidth]{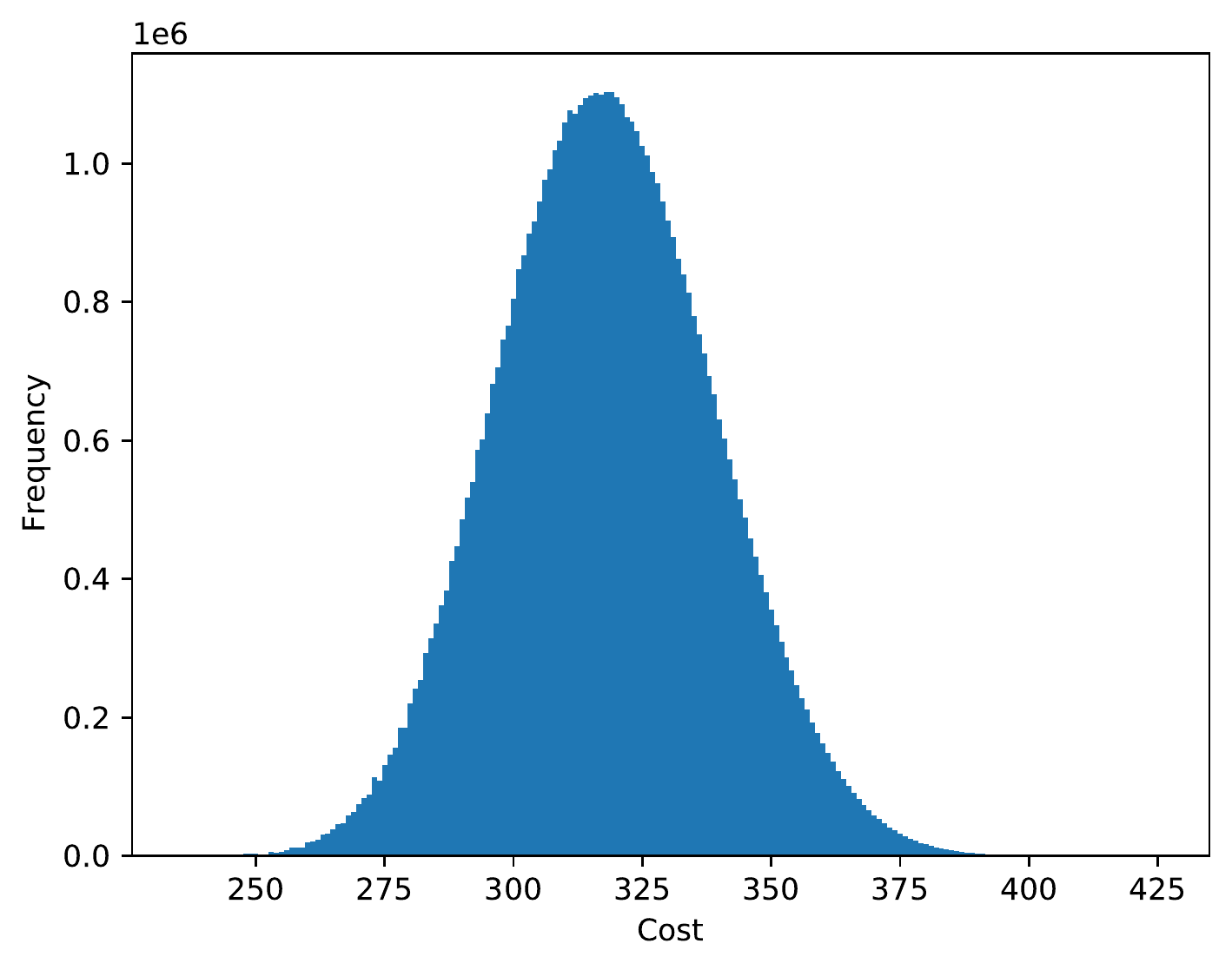}
            \caption{}
            \label{fig:CVRP_linear}
        \end{subfigure}
        \begin{subfigure}{1\columnwidth}
            \centering
            \includegraphics[width=0.85\columnwidth]{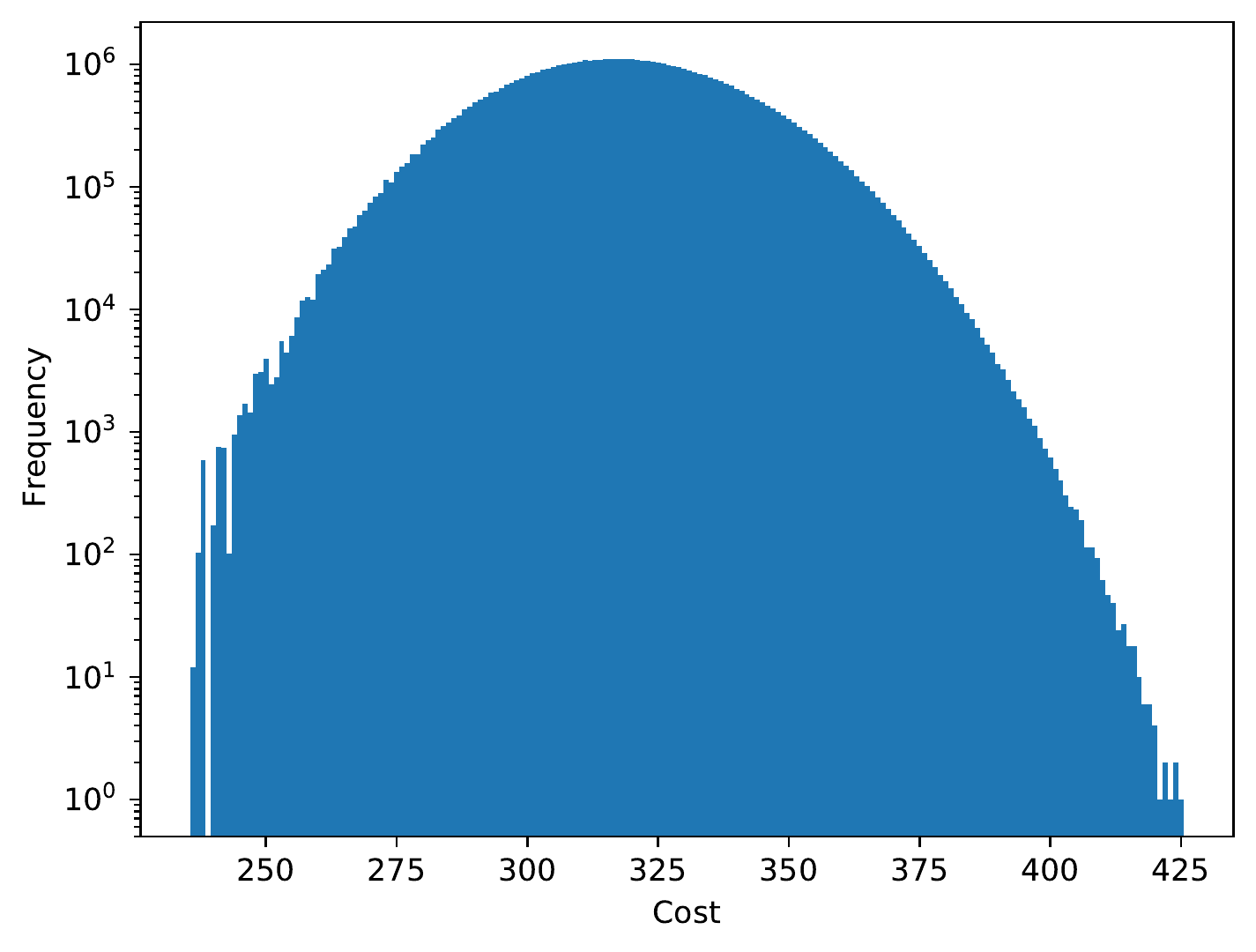}
            \caption{}
            \label{fig:CVRP_log}
        \end{subfigure}
        \caption{Distribution of solution qualities for the randomly generated 10-location vehicle routing problem,  shown with (a) a linear scale and (b) a log scale.}
        \label{fig:CVRP_distribution}
    \end{figure}

    To generate a quality distribution, the package vector was randomly generated from integers on the interval [5,30], a symmetric cost matrix was generated with depot to location costs randomly generated from integers [10,20] and inter-location costs from integers [1,15], and finally, the vehicle capacity was taken as 20. Note that integer values were used in the cost matrix to increase degeneracy in the quality distribution of the solution space, just to provide some variety compared to the portfolio optimisation problem which shows virtually no degeneracy. The distribution of solution qualities is shown in \cref{fig:CVRP_linear}, where it is clear that the qualities are distributed in what approximates a normal distribution. The same distribution is shown with a log-scale in \cref{fig:CVRP_log}, in which it becomes more clear that the distribution is not perfectly normally distributed, but rather, the distribution at lower costs is somewhat discontinuous and truncated. In fact, there are 12 solutions which share the lowest cost.
    
\begin{figure}[H]
        \centering
        \includegraphics[width=0.95\columnwidth]{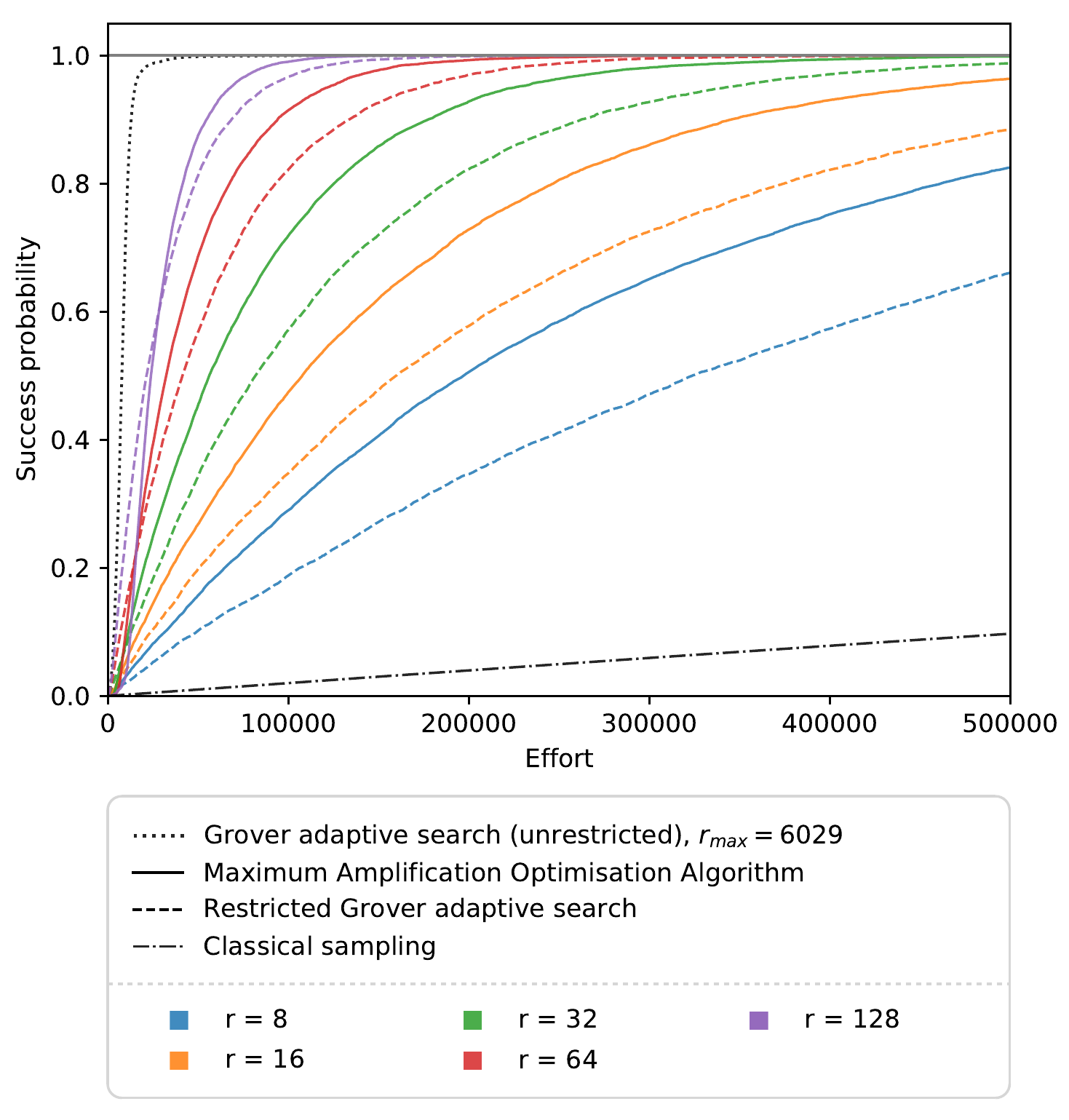}
        \caption{Simulation results for a large vehicle routing problem. The MAOA consistently outperforms the RGAS, both of which significantly outperform a classical random-sampling approach.}
        \label{fig:CVRP_results}
    \end{figure}

    The simulation results of the MAOA, GAS and RGAS applied to this 10-location CVRP are shown in \cref{fig:CVRP_results} along with the classical random-sampling method, where the success probability refers to the probability of one of the 12 highest quality (lowest cost) solutions being measured. 
    The GAS method is included to show how an algorithm unrestricted in circuit depth would perform. The rotation count that would be required to produce complete convergence into a single optimal solution over a solution space of this size is given by $r_c=6,029$. Since a user knows the size of the solution space, but not the degeneracy of the optimal solution(s), this is the user-specified maximum rotation count required for the GAS in this instance. In contrast, the MAOA and the RGAS are tested with restricted circuit depths corresponding to rotation counts of $r \in \{8,16,32,64,128\}$. In all cases, the quantum algorithms provide speedup over the classical method, but the amount of speedup increases with increasing rotation counts. The MAOA also outperforms the RGAS, as predicted in \cref{sec:GAS}. To clarify, due to the higher upfront computational expense of the threshold finding process, the RGAS begins to perform better than the MAOA as $r$ approaches $r_c$, but this is not likely to be relevant in the context of large solution spaces and restricted circuit depth near-term quantum computing.
    
     \begin{figure*}
        \centering
        \includegraphics[width=1.3\columnwidth]{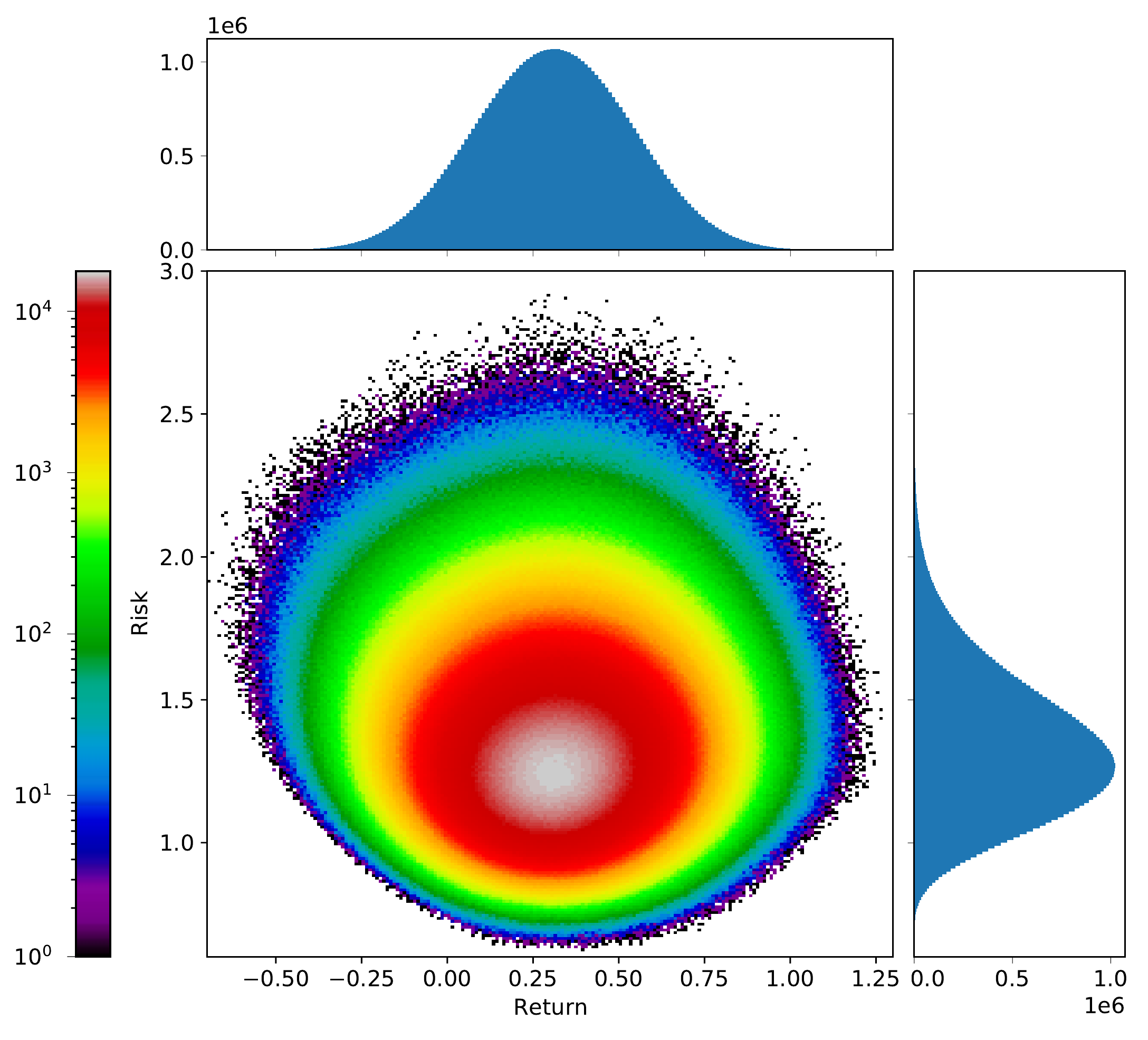}
        \caption{Distribution of expected returns and risks associated with each portfolio choice for a large portfolio optimisation problem.}
        \label{fig:Portfolio_distribution}
    \end{figure*}
    
    \subsection{Portfolio optimisation}
    
    It was demonstrated by \citet{portfolio} that a portfolio optimisation problem based on the Markowitz model \cite{markowitz_portfolio} was a suitable candidate for application of the QWOA framework, as such, it makes an equally suitable candidate for the MAOA and GAS methods. This work will give a somewhat different treatment to the problem, however, by treating the two components of the objective function, risk and expected return, seperately. Given $n$ different stocks/assets, a particular choice of portfolio can be expressed by $\bm{z}=(z_1,z_2,...,z_n)$, where for each asset $i$ we have $z_i \in \{-1,0,1\}$, where each value corresponds with a short position, no position, or long position, respectively. In addition, the portfolio positions are constrained by the net position, $I$, such that:
     \begin{equation}
         \sum_{i=1}^{n}{z_i}=I.
     \end{equation} 
    
    Under the Markowitz model, using data for the daily expected percentage return of asset $i$, $R_i$, and the co-variance values between assets $i$ and $j$, $\sigma_{ij}$, for a group of $n$ assets, the expected return and associated risk for each portfolio choice can then be characterised by:
    \begin{equation}
        \label{eq:return}
        \text{Return} = \sum_{i=1}^{n}{R_i z_i}
    \end{equation}
    \begin{equation}
        \label{eq:risk}
        \text{Risk} = \sum_{i,j=1}^{n}{\sigma_{ij} z_i z_j}.
    \end{equation}
    
    The number of unique and valid portfolio choices available, or the cardinality of the solution space, is given by:
    \begin{equation}
        N_P(n,I) = \sum_{s=0}^{\left \lfloor{\frac{n-I}{2}}\right \rfloor}{{n \choose I+s}{n-I-s \choose s}},
    \end{equation}
    where $s$ represents the number of possible shorts, the first term represents the placement of longs within $\bm{z}$, and the second term represents the subsequent placement of the shorts. For the purpose of generating an example solution space and corresponding quality distribution, data for the daily adjusted close prices from 01/01/2019 to 31/12/2020 was analysed for 20 different stocks from the ASX.20 index: AMP, ANZ, AMC, BHP, BXB, CBA, CSL, IAG, MQG, GMG, NAB, RIO, SCG, S32, TLS, WES, BKL, CMW, HUB, ALU.
    
    The net position, $I$, was taken to be 7, such that the total number of solutions or portfolio choices was $N(20,7)=61,757,600$. The distribution of risks and returns for the resulting solution space is shown in \cref{fig:Portfolio_distribution}, where risks are scaled down by a factor of $100$. The distribution resembles a 2D Gaussian, skewed towards high risk portfolios. In order to navigate the 2-dimensional optimisation landscape, the marking function will use two thresholds, one for risk and one for return. It is presumed that a balance between optimising for low risk while still maximising returns is desirable. As such, the risk threshold will be set and fixed at that corresponding to the lowest $10\%$ of all solutions, transforming the optimisation problem into a maximisation of expected return within the subspace of low-risk solutions. As is discussed in more detail in \cref{sec:MAOA_benefits}, the MAOA presents a unique ability to be able to navigate multidimensional optimisation landscapes, but because the GAS/RGAS do not have the same ability, the problem must be transformed into a 1 dimensional problem to allow for effective comparison between the different methods. 
    
\begin{figure}[ht]
        \centering
        \includegraphics[width=0.95\columnwidth]{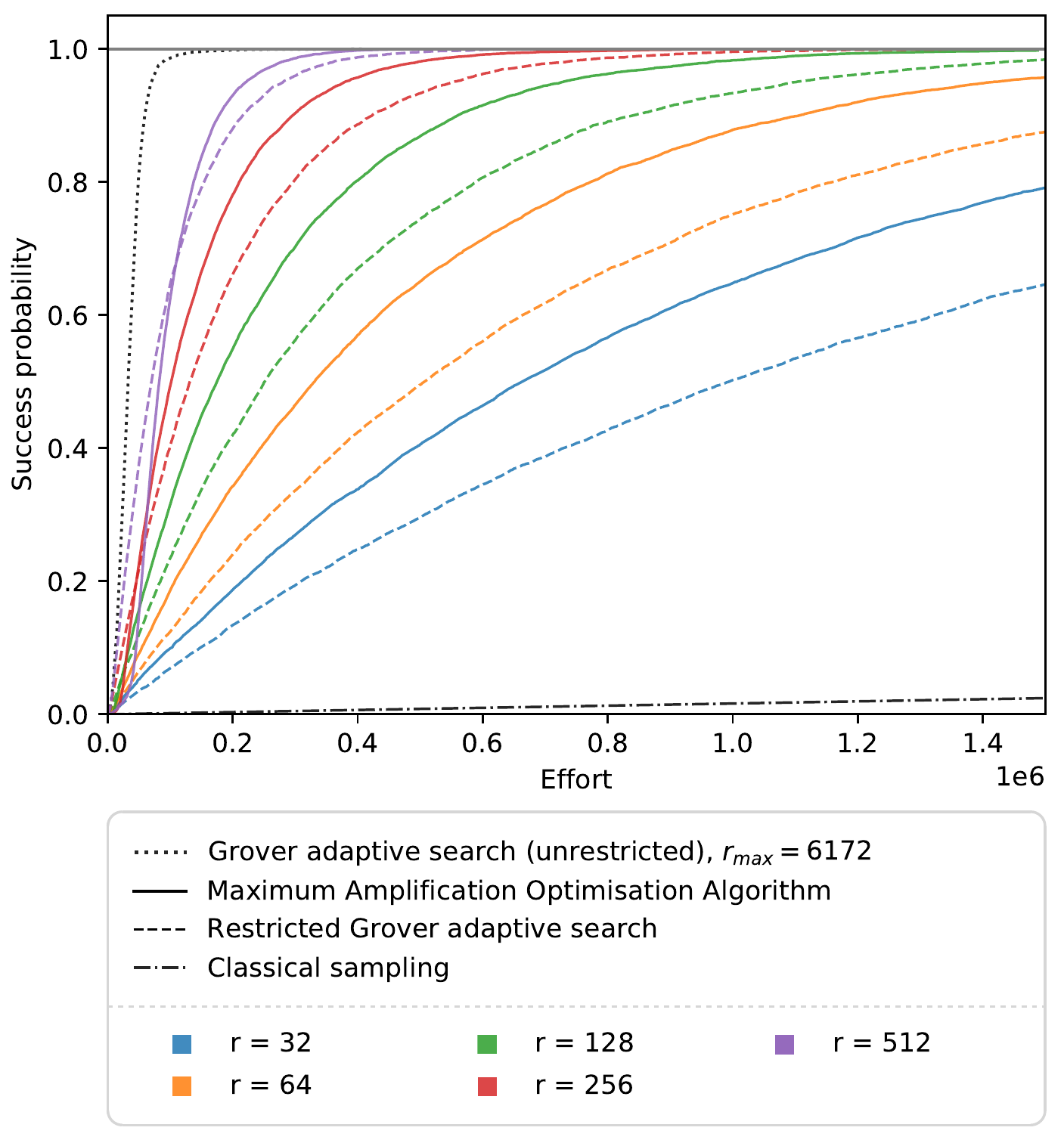}
        \caption{Simulation results for a large portfolio optimisation problem. The MAOA consistently outperforms the RGAS, both of which significantly outperform a classical random-sampling approach.}
        \label{fig:Portfolio_results}
    \end{figure}

    The simulation results for this problem are shown in \cref{fig:Portfolio_results}, where the probability of success is taken as the probability of finding the single highest return portfolio from those within the lowest $10\%$ for risk. The user-specified maximum rotation count required for the GAS in this instance is given by $r_c=6,172$, which is derived directly from the known size of the solution space. In contrast, the MAOA and the RGAS are tested with restricted circuit depths corresponding to rotation counts of $r \in \{32,64,128,256,512\}$. The reason these rotation counts are higher than for the CVRP problem, is because here, there is only a single optimal solution, compared to 12 in the case of the CVRP problem. The smaller ratio of optimal solutions requires higher rotation counts for comparable performance. In any case, the results are consistent with those for the CVRP simulations. In all cases, the quantum algorithms provide speedup over the classical method, with the amount of speedup increasing with increasing rotation counts. The MAOA also once again consistently outperforms the RGAS.
    
    \subsection{Simulating an arbitrarily large problem}
    
    The above problems are of a scale where the quality distributions can be readily computed on a desktop computer. In reality, the MAOA is designed to be applied to problems with significantly larger solution spaces, which are intractable through classical methods. It is therefore valuable to understand how the MAOA performs in the limit of very large problems. As can be seen in the CVRP and portfolio optimisation problems, combinatorial optimisation problems often possess solution spaces which have qualities that are normally distributed. It is therefore possible to simulate an arbitrarily large problem by using the standard normal distribution. It is not unreasonable to think that a large enough problem would have a quality distribution which is normally distributed and which approximates a continuous distribution, even at large deviations from the mean. For such a problem, it is therefore possible to take the marked-solution ratio, $\rho(T)$, as the cumulative distribution function of the standard normal distribution at $T$. This allows for the RGAS and MAOA to be simulated in optimising an arbitrarily large normally distributed problem. 
    
    For the purpose of assessing the behaviour of the RGAS and MAOA algorithms in the limit of large problems, they will be analysed for a constant restricted rotation count, $r=64$. In each case, they will be seeking a solution within a certain most-optimal target group, forming a fraction of the solution space, referred to as the target ratio, $\mu$. A smaller target ratio corresponds with a search for higher quality solutions. The RGAS finds the target high-quality solutions by sequentially partitioning the solution space into smaller and smaller improving subsets, until a solution within this target group is measured. On the other hand, for a maximum rotation count of $r=64$, the MAOA repeatedly measures from a state prepared with a threshold which produces maximum amplification. Since this is known to occur when the probability of successfully measuring a marked solution, $P(r,T) \leq \frac{1}{40}$, and also when the probability is amplified by a factor of $(2r+1)^2$, the final marked-solution ratio will be approximated by: 
    \begin{equation}
        \label{eq:final_marked_ratio}
        \rho(r)=\frac{1}{40(2r+1)^2}.
    \end{equation}
    
\begin{figure}[H]
        \centering
        \includegraphics[width=0.95\columnwidth]{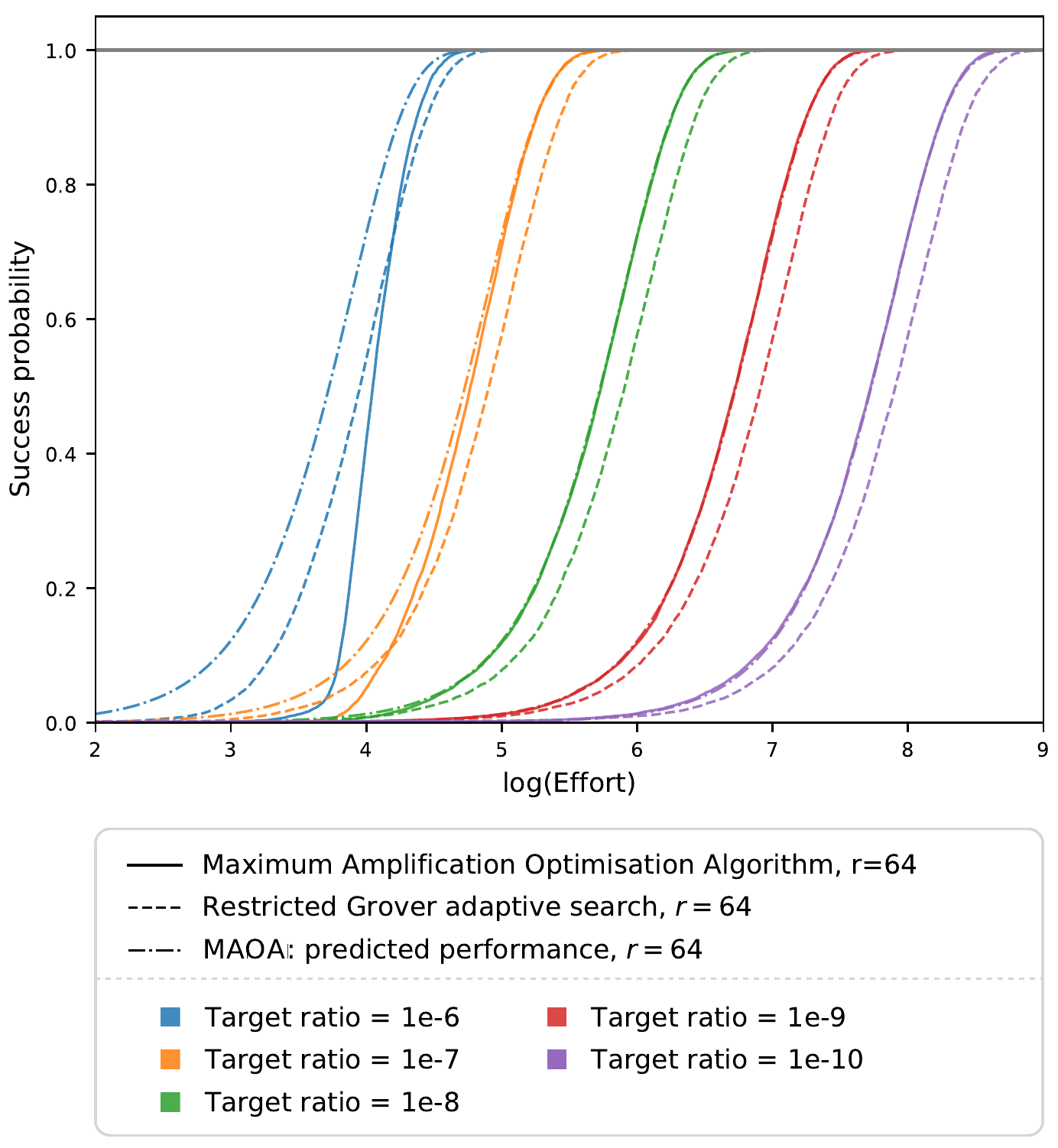}
        \caption{The simulated performance of both the MAOA and RGAS in optimising arbitrarily large, normally distributed solution spaces. Theoretically predicted curves for the MAOA are also included, which are derived in \cref{sec:MAOA_analysis}. The MAOA consistently outperforms the RGAS when searching for the highest-quality solutions.}
        \label{fig:SND_results}
    \end{figure}

    So in this case, the MAOA will be somewhat regularly measuring marked solutions within roughly the top $\rho=\num{1.5e-6}$, regardless of the target ratio, $\mu$. It is through this method that the MAOA seeks to find a solution from the target group. The performance of each algorithm has been simulated over a range of target ratios, $\mu \in \{10^{-6},10^{-7},10^{-8},10^{-9},10^{-10}\}$, and the results are shown in \cref{fig:SND_results}. The MAOA is shown to perform better than RGAS in the limit of small $\mu$, or in other words, the MAOA consistently finds solutions of the highest qualities faster than RGAS.
    
    \section{Analysis of the Maximum Amplification Optimisation Algorithm in the large problem limit}
    \label{sec:MAOA_analysis}
    
    To understand how well the MAOA performs relative to a classical random-sampling of the solution space, it is useful first to quantify the probability of successfully finding a target solution for the classical case. Since the behaviour of interest is that in the limit of very large solution spaces, it is suitable to ignore the removal of sampled solutions from the solution space (i.e. sampling with repeats). Note this assumption is only reasonable when the target ratio is much larger than that for a single target solution, $\mu\gg\frac{1}{N}$. Given this assumption, the equation for the probability of success in the classical case is given by:
    \begin{equation}
        P_C(e,\mu)=1-(1-\mu)^{e}.
    \end{equation}
    
    Note that $e$ refers to the computational effort, but in this case, can also be understood as the number of classical samples taken, since they are both equivalent. The equation is best understood as being derived from the complement of a successful measurement, i.e. the probability of success after $e$ samples is equal to one subtract the probability of no successes after $e$ samples. The probability of failing to sample a target solution $e$ times in a row is clearly $(1-\mu)^e$, since $\mu$ is the probability of a successful sample.
    
    The equation for probability of success for the MAOA can be derived in a similar fashion, Since the MAOA, once an appropriate threshold has been found, essentially reduces to repeated preparation and subsequent measurement of the maximally amplified state. The equation is therefore given by:
    \begin{equation}
        \label{eq:success_prob_MAOA}
        P_Q(e,\mu,r)=1-(1-\mu (2r+1)^2)^{\frac{e}{2r+1}}.
    \end{equation}
    
    The $(2r+1)^2$ term relates directly to the maximum amplification of marked states due to $r$ rotations. The power, $\frac{e}{2r+1}$, gives the number of measurements, since each state preparation and measurement requires computational effort $(2r+1)$. As can be seen in \cref{fig:SND_results}, this analytically derived expression for the success probability is consistent with simulation results in the limit of small target ratios (i.e. those significantly smaller than the final value for $\rho$).
    
    Rearranging each of these equations for the effort in the classical case, $e_C$, and the quantum case, $e_Q$, taking their ratio and taking the limit in small target ratios, an expression for the speedup of the MAOA over classical random-sampling can be derived:
    \begin{equation}
        \lim_{\mu\to0} \frac{e_C}{e_Q}=\lim_{\mu\to0} \frac{\log(1-\mu(2r+1)^2)}{(2r+1)\log(1-\mu)}=2r+1.
    \end{equation}
    
    Note that this result can also be understood intuitively, since each state preparation provides amplification of $(2r+1)^2$ at the expense of $(2r+1)$ computational effort, what remains is a speedup of $(2r+1)$. This result essentially implies that the MAOA is capable of producing speedup (over classical random-sampling) in finding near-optimal or optimal solutions to large combinatorial optimisation problems, and that this speedup grows linearly in the achievable circuit depth. Since the MAOA produces states in which optimal solutions are maximally amplified, and does so without a computationally expensive variational procedure, it represents the upper limit of speedup available in the context of restricted circuit depths using a deterministic quantum amplitude amplification protocol.
    
    It is worth noting, however, that the maximum amplification of optimal solutions produced by the MAOA at a given restricted circuit depth is maximum within the class of ``amplitude amplification" algorithms. That is, within the class of algorithms which amplify target states through the interleaved application of phase shifts and diffusion/mixing operators. Examples include the rotation and diffusion operators of Grover's algorithm, as well as the alternating operator ansatz of QAOA/QWOA. It may be possible to project into high quality solution states with significantly less circuit depth by using circuit ansatz outside of the ``amplitude amplification" framework. One such example is the filtering variational quantum algorithm by \citet{amaro2022filtering}, which has shown promise to converge to optimal solutions in significantly fewer operations compared with QAOA for a MaxCut problem on random cubic weighted graphs. However, an overall speedup is yet to be demonstrated for such methods, due to the computational expense of the required variational procedures. 
    
    \label{sec:MAOA_benefits}
    The MAOA presents additional useful features beyond its demonstrated speedup:
    
    \begin{enumerate}
        \item The analytically derived expression in \cref{eq:success_prob_MAOA} can be used to inform a user how likely they are to have found a solution within a particular target ratio, $\mu$, after a specified amount of computational expense.
        \item The MAOA also provides additional flexibility for multi-dimensional optimisation problems. For example, in the portfolio problem, the peak finding process, implemented over the risk threshold, can be used to isolate a known fraction of the lowest risk options. Fixing the risk threshold and transitioning to optimisation over the return threshold then allows the user to optimise within this space of lowest risk options. Note that this multi-stage optimisation procedure can be generalised to other problems too.
        \item Repeated sampling of the MAOA amplified state produces a large set of near-optimal solutions. In contrast, the RGAS only produces one solution at each of the measured improving qualities. At the tail end of the RGAS procedure, these near optimal/improving solutions would be measured extremely rarely. On the other hand, for the MAOA, the near-optimal solutions are measured regularly throughout the duration of sampling. Note that the ratio of these regularly sampled marked solutions can be approximated as per \cref{eq:final_marked_ratio}. This is beneficial in acquiring a significantly larger group of near-optimal solutions, which may be of interest in some cases. For example, it may allow one to then select between solutions for features not accounted for within the optimisation procedure or alternatively to simply have access to back-up high quality solutions.
    \end{enumerate}
    
\section{Conclusion}
\label{sec:Conclusion}

    This paper serves as a comprehensive introduction to the Maximum Amplification Optimisation Algorithm (MAOA), a near-term quantum algorithm designed for finding high quality solutions to large combinatorial optimisation problems, while constrained to restricted circuit depths. Other existing near-term algorithms, QAOA and QWOA, focus on producing amplified states in which the expected value of quality has been optimised. When measuring from such an amplified state, we expect to find a solution of high quality. However, as we have demonstrated in this paper, the highest quality solutions in such a state are not amplified to the maximum extent possible. 
    
    The MAOA shifts the paradigm by seeking an amplified state in which the highest quality solutions are maximally amplified, then repeatedly sampling from the maximally amplified state. Since the highest quality solutions are amplified to the maximum extent possible, subject to a given circuit depth, the frequency with which they will be measured is also maximised, hence delivering maximum possible speedup. 
     
    Perhaps more importantly, we demonstrate that these maximally amplified states can be produced via a known set of parameters, which removes entirely the computationally expensive variational process typically associated with the QAOA and QWOA algorithms. As such, we demonstrate that the MAOA is capable of producing optimal solutions to large combinatorial optimisation problems faster than through classical exhaustive search, a result which has remained elusive for variational algorithms.
    
    
    \section{Acknowledgements}
    
    We would like to thank Sam Marsh and Edric Matwiejew for their important contribution to this work. To Sam, thank you for providing a number of important insights, including inspiration for the graph reduction process, as well as pointing out the connections between QWOA and Grover's search. To Edric, thank you for all your work developing and assisting in the use of the QWOA simulation software with which the proficiency of a binary marking function was first noticed, leading the way to the MAOA's development. This work was also supported with resources provided by the Pawsey Supercomputing Centre, with funding from the Australian Government and the Government of Western Australia. 
    
	\bibliography{refs}
	\clearpage
	\appendix
	\section{Investigation of amplification of a single vertex}
    \label{sec:AppendixA}
    
    In order to understand how a graph's structure may effect its capacity to produce amplification at a single vertex, it is important to first define some parameters. The first is the average degree of a graph, which is the average number of edges connected to each vertex on the graph, referred to from here on by the variable, $D$. The second parameter considered, relates to the spectral quality of the graph's adjacency matrix, or more specifically, the number of distinct eigenvalues possessed by the graph's adjacency matrix, from here on referred to as spectral count, $E$. In referring to different graphs within the following figures, the parameters will be combined into a single label: $DxEy$, where $x$ is the value for $D$, and $y$ is the value for $E$.
    
    In order to assess how amplification varies with these parameters, a number of 24-vertex circulant and connected graphs ($N=24$) have been analysed. The reason for choosing 24 vertices in particular is because of its large number of divisors, allowing for a large number of unique circulant graph structures.  Each graph has been characterised by an adjacency matrix consistent with some specific values for $D$ and $E$. Three iterations of the QWOA process were applied to each graph to generate the final state of each graph, $\ket{\bm{\gamma}, \bm{t}}$, given by \cref{eq:QWOA} in combination with 6 variational parameters, $\bm{t} = (t_1, t_2, t_3)$ and $\bm{\gamma} = (\gamma_1, \gamma_2, \gamma_3)$. Each graph was assigned 48 different randomly generated quality distributions with each quality randomly sampled from a uniform distribution over $[0,1)$. The 6 parameters were optimised in each case to produce maximum amplification in the optimal vertex. The optimisation procedure consists of producing 10,000 randomly generated sets of initial parameters. Each of the 10 parameter sets that produce maximum initial amplification is used as the initial set of values in a Nelder-Mead optimisation procedure. The largest final probability produced from these 10 optimisation procedures is taken to be the final amplified probability for that particular graph and quality distribution. This optimisation procedure was repeated for the 48 different quality distributions assigned to each graph, from which the mean and standard deviation of the amplified probability was computed.
    
\begin{figure}[ht]
        \centering
        \begin{subfigure}{1\columnwidth}
            \centering
            \includegraphics[width=0.95\columnwidth]{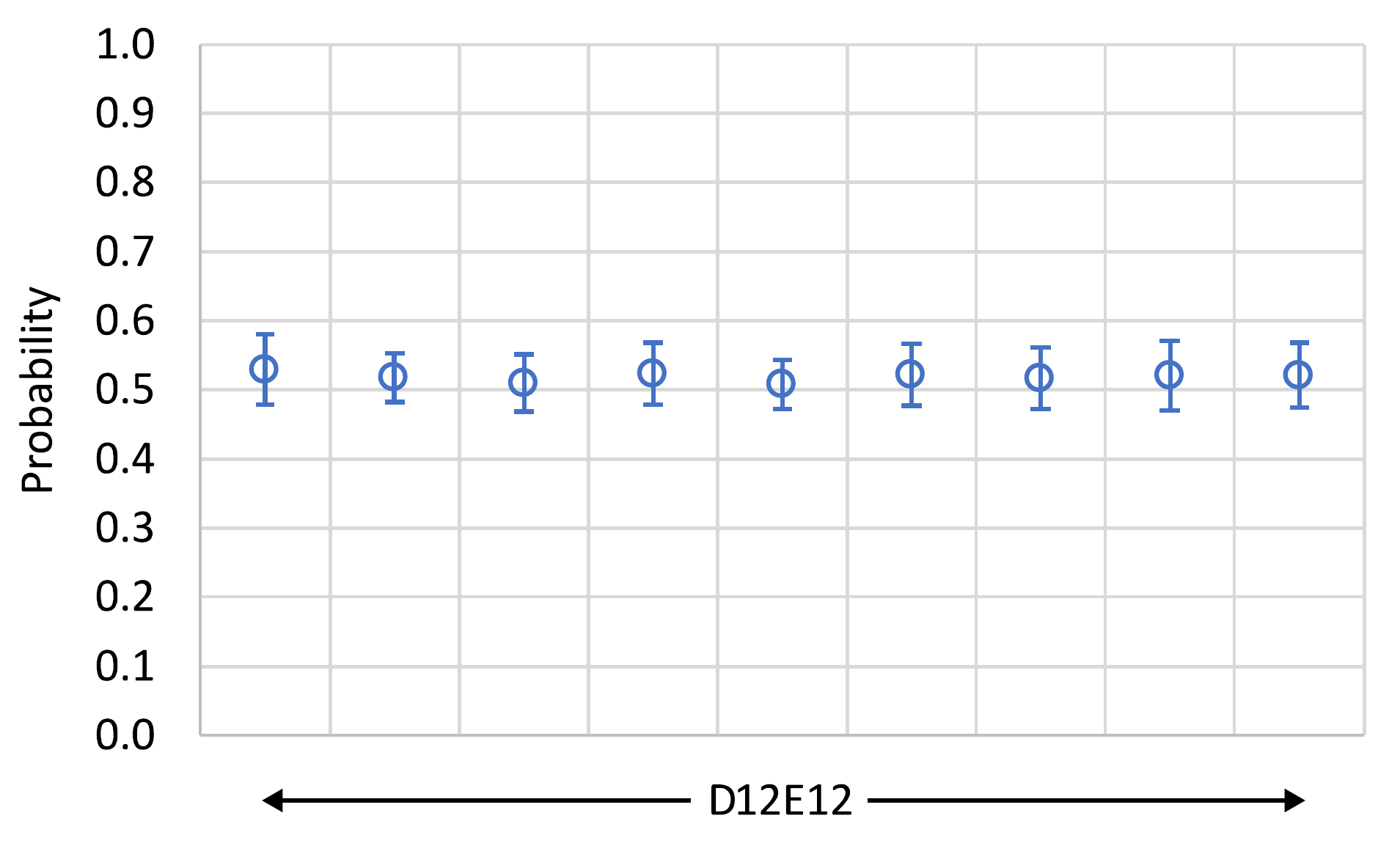}
            \caption{}
            \label{fig:constant_structure}
        \end{subfigure}
        \begin{subfigure}{1\columnwidth}
            \centering
            \includegraphics[width=0.95\columnwidth]{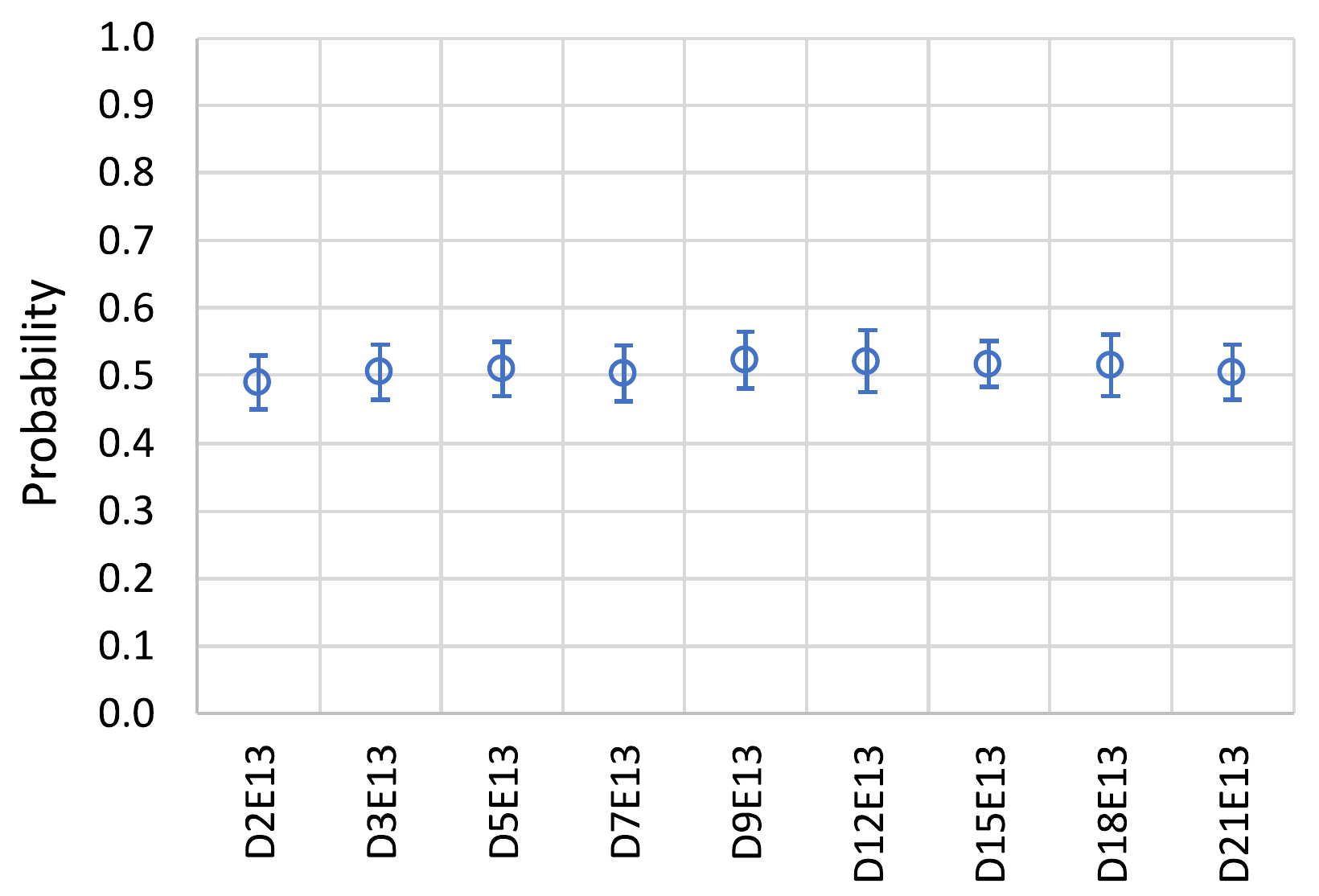}
            \caption{}
        \label{fig:varying_degree}
        \end{subfigure}
        \begin{subfigure}{1\columnwidth}
            \includegraphics[width=0.95\columnwidth]{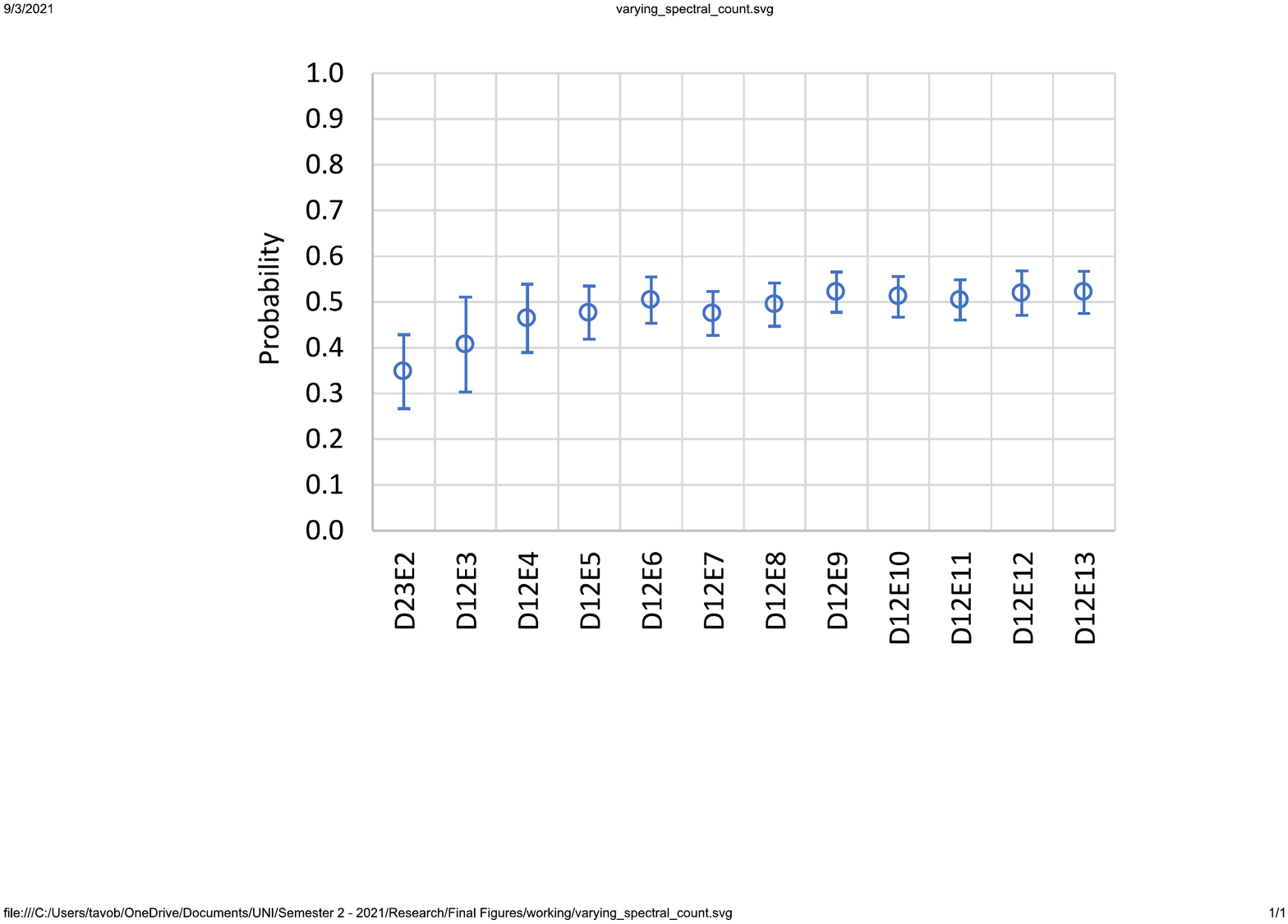}
            \caption{}
            \label{fig:varying_E}
        \end{subfigure}
        \caption{Amplified probabilities of a single optimal vertex for graphs with varying spectral count and average degree.}
    \end{figure}

    Perhaps the first thing to clarify, is that graphs with consistent values for $D$ and $E$ produce consistent amplification, to ensure there is not some other important factor requiring consideration. As such, 9 distinct graphs, each with values $D=12$ and $E=12$, were generated randomly from all such circulant graphs. The final amplified probabilities for each of these 9 graphs is shown in \cref{fig:constant_structure}, from which it can be concluded that there is no significant difference in amplification between them. As such, it seems likely that circulant graphs with the same average degree and spectral count produce consistent behaviour in terms of amplification of a single optimal vertex.
    
    It may be natural to suspect a higher average degree might increase the rate at which a single optimal node is amplified, but in fact, it appears to make very little difference, as shown in \cref{fig:varying_degree}, for which the plot was produced by fixing the spectral count at 13, and randomly selecting graphs with increasing average degrees ranging from $D=2$ to $D=21$. This may be because even with low average degrees, continuous time quantum walks are able to mix across arbitrarily large distances between vertices. 
    
    On the other hand, spectral counts do appear to have a noticeable effect on a graph's ability to produce amplification of a single optimal vertex. By fixing the average degree at 12, and selecting random graphs with increasing spectral counts ranging from $E=3$ to $E=13$, the plot in \cref{fig:varying_E} was produced. Note that an additional graph, $D23E2$, which is the complete graph, $K_{24}$, was included because this is the only graph with only 2 distinct eigenvalues (a fact which is consistent for graphs with any number of vertices). From these results, it appears that for randomly distributed quality distributions, graphs with smaller spectral counts are less effective at amplifying a single optimal vertex.
    
\begin{figure}[ht]
        \centering
        \includegraphics[width=0.95\columnwidth]{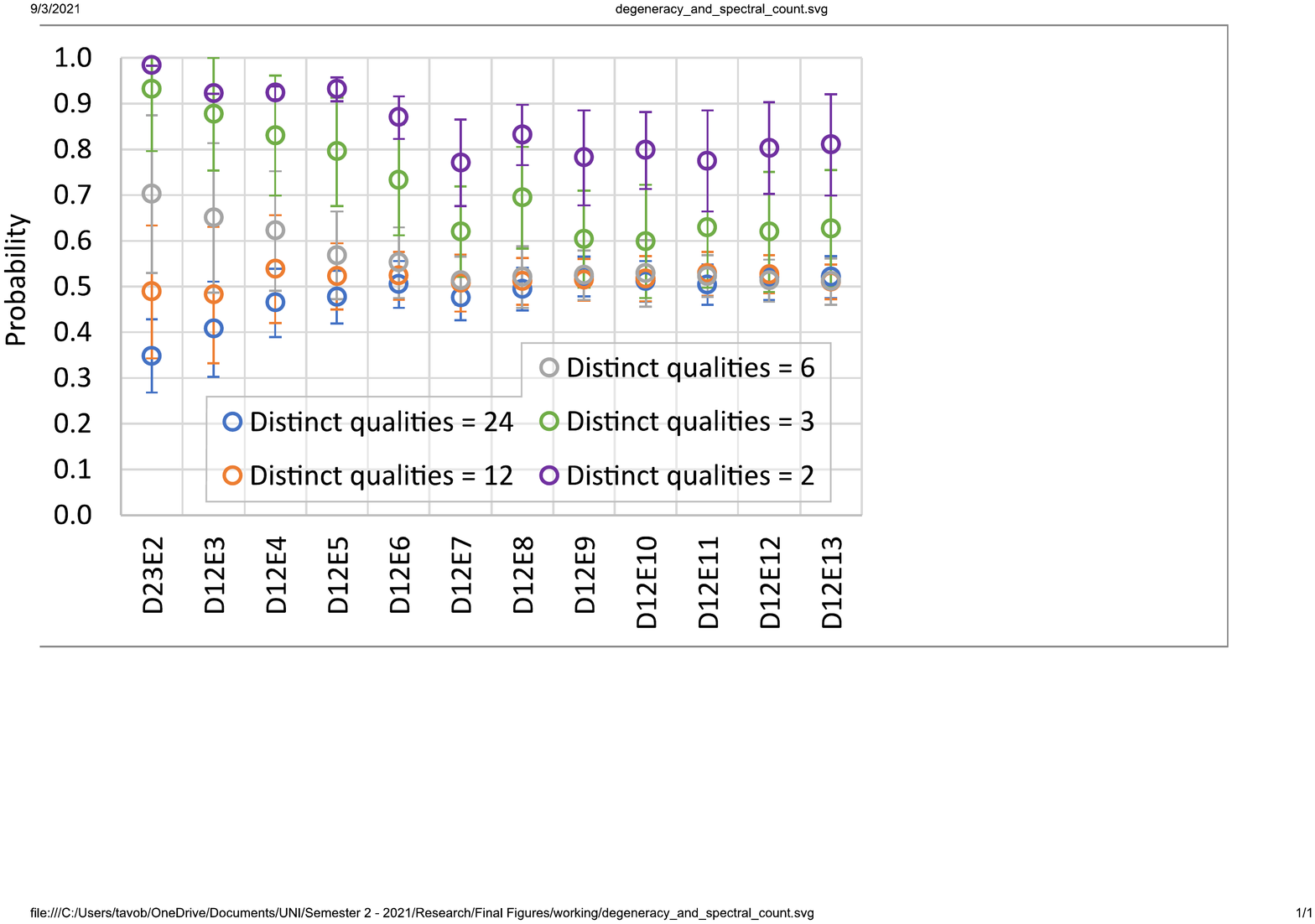}
        \caption{Amplified probabilities of a single optimal vertex for graphs of varying levels of degeneracy in the quality distributions.}
        \label{fig:degen_and_E}
    \end{figure}

    Referring to \cref{fig:degen_and_E}, it turns out that this dependence of amplification on spectral count is actually highly sensitive to degeneracy within the quality distribution, with the relationship reversing for cases with higher degeneracy, where graphs with lower spectral counts produce higher amplifications. Further to this, and perhaps most importantly of all, degeneracy in the quality distribution appears to be the single most critical feature in terms of influencing optimal node amplification for graphs in general, with higher degeneracy allowing for higher amplification. Note that the analysis for varying levels of degeneracy was completed in the same way as for the no degeneracy case, except that the 24 qualities were generated by repetition of values from a smaller set of randomly generated values, assigning the optimal quality to only one vertex, and randomising the arrangement of the final set of qualities across the 24 vertices.
    
    Taking all of this together, it appears that the combination of quality distributions with high degeneracy on graphs with low spectral count provide the highest amplification into a single marked vertex. The most effective amplification therefore appears to occur with a binary marking function over a complete graph, at least for the case of a single optimal vertex.
    
    \section{Investigation of the optimisation landscape}
    \label{sec:AppendixB}
    Under the QWOA framework it's also important to consider how a graph's structure and quality distribution effect the ruggedness of its optimisation landscape, and the variability in its local extrema. It may still be possible that a graph which produces lower amplification than another, still might be superior if an optimal set of parameters can be found with sufficiently less computational effort.
    
\begin{figure}[H]
        \centering
        \includegraphics[width=0.95\columnwidth]{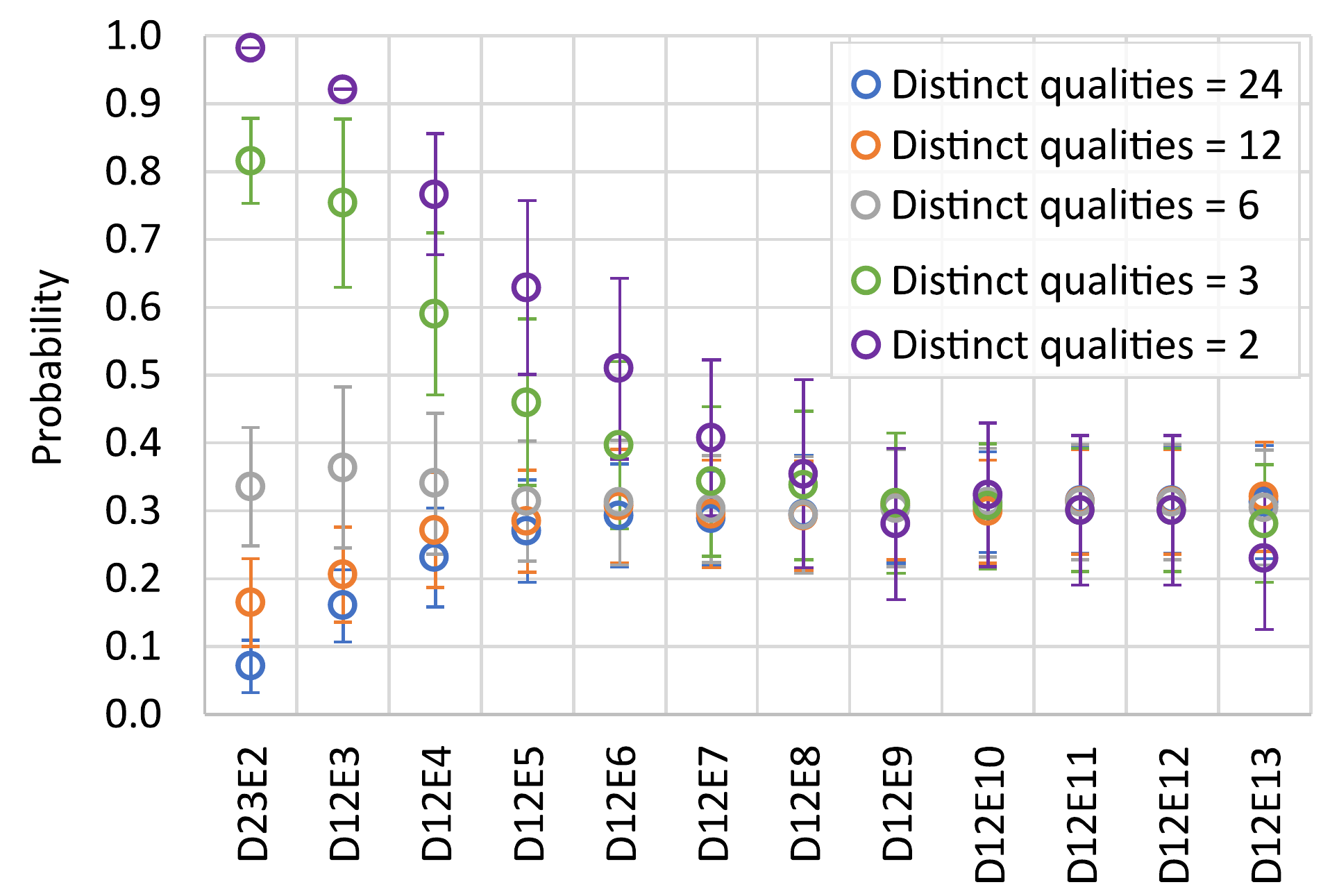}
        \caption{Amplified probabilities of a single marked vertex for varying levels of degeneracy in the quality distributions.}
        \label{fig:optimisation_landscape}
    \end{figure}

    In order to assess the optimisation landscapes, each graph in a series of graphs with varying spectral counts is assigned the same uniformly distributed qualities over the interval $[0,1]$, with a single marked vertex assigned a quality of 1. This is repeated for quality distributions with varying levels of degeneracy. In each case, 240 initial parameter sets are generated and taken as initial values is a Nelder-Mead optimisation procedure. From the 240 optimisation results, we take the mean and standard deviation for the maximised probability in the single marked vertex. As such, the mean value is a measure of the average quality (defined by the amplified probability) of local maxima, and the standard deviation is a measure of how much local maxima vary in quality across the optimisation landscape. The results of this procedure are shown in \cref{fig:optimisation_landscape}. 
    
    The general result is that high degeneracy quality distributions combined with graphs of low spectral count show the best performance in terms of the average amplified probability produced at each local maxima. Even though they still have comparable variability in local maxima quality, this would make the optimisation procedure easier in terms of finding a ``good enough" solution. A more specific result is that all 240 initial parameter sets converge to the same peak amplified probability for the binary E=2 and E=3 cases, as shown by the zero offset error bars. As we already demonstrate, binary marking on the complete graph produces the best amplification, but it also appears to do so with an optimisation landscape in which the local maxima are all equally optimal, meaning that an optimisation procedure for this case is likely to be highly efficient. It's worth noting the wide ranging performance of the complete graph ($D23E2$) and the other low spectral count graphs. They perform poorly in the general case (small amount of degeneracy) but exceedingly well in the binary case, and in varying degree between these two extremes for the intermediate cases.
    
\begin{figure}[H]
        \centering
        \begin{subfigure}{1\columnwidth}
            \centering
            \includegraphics[width=0.95\columnwidth]{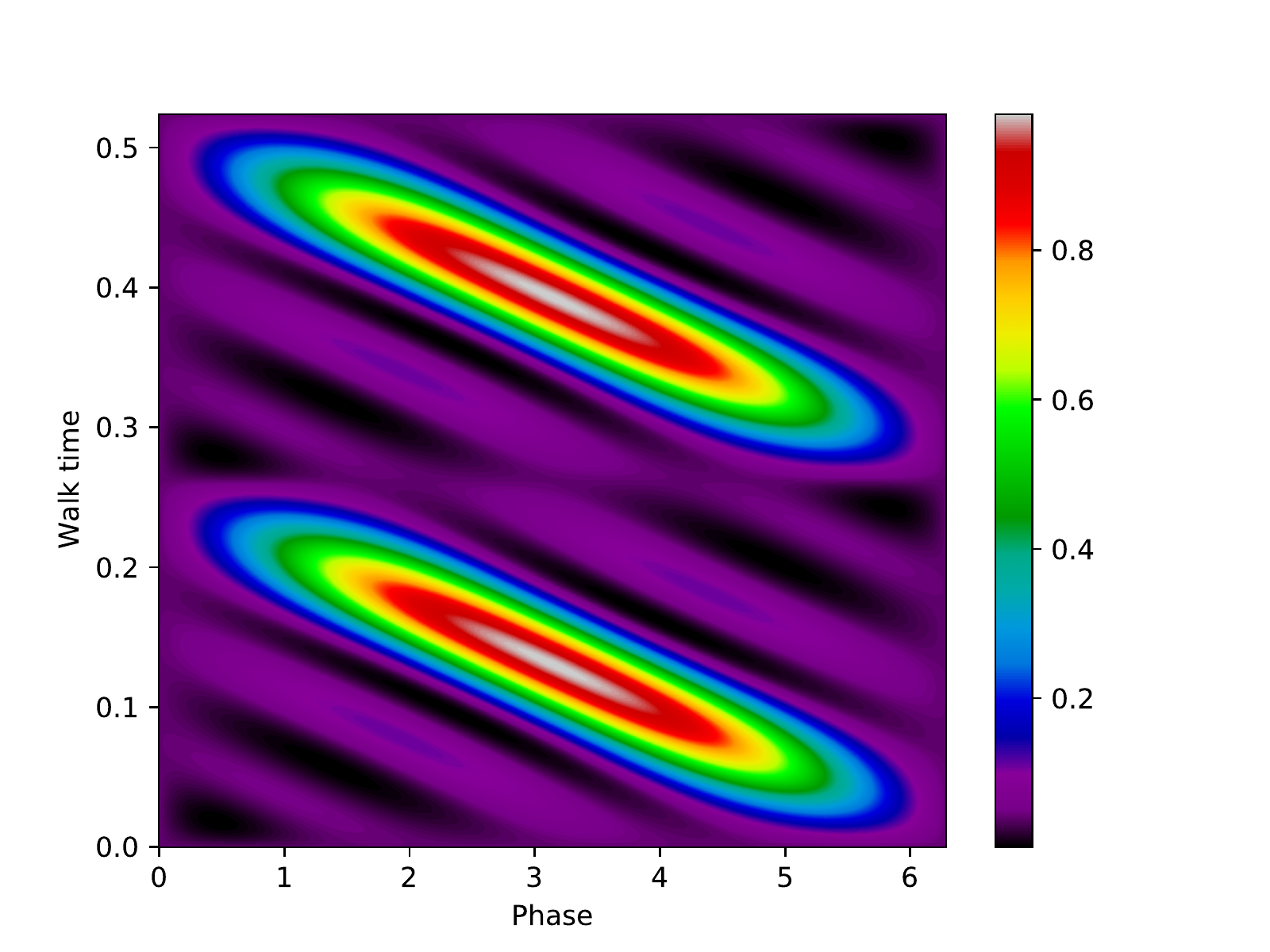}
            \caption{}
            \label{fig:fixed_pairs_binary}
        \end{subfigure}
        \begin{subfigure}{1\columnwidth}
            \centering
            \includegraphics[width=0.95\columnwidth]{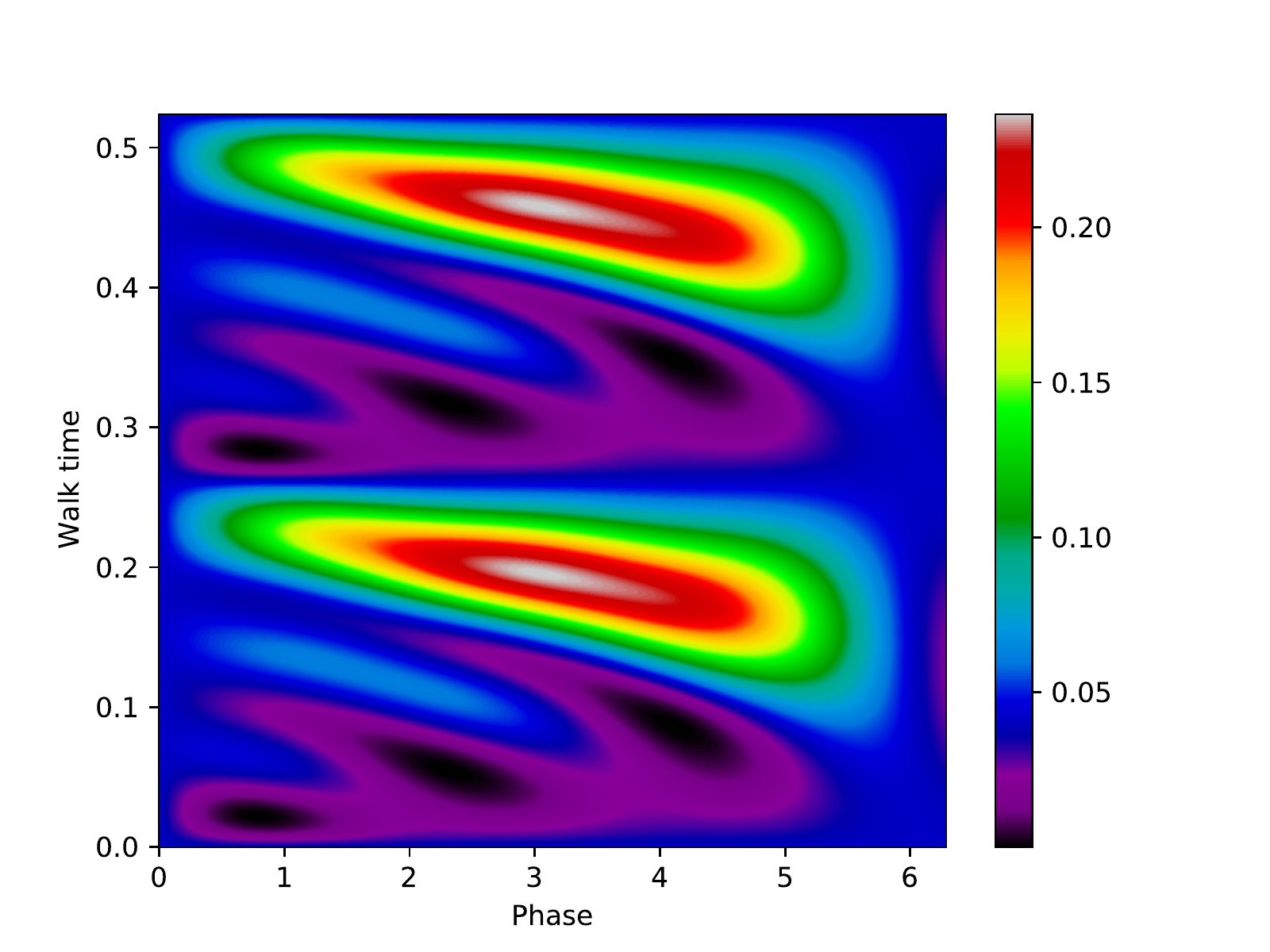}
            \caption{}
            \label{fig:fixed_pairs_nonbinary}
        \end{subfigure}
        \caption{Maximum amplified probability achieved with repeated parameters is significantly larger for (a) the binary quality distribution ($P_{max}=0.98$) compared with (b) the uniform distribution of qualities ($P_{max}=0.24$).}
        \label{fig:quality_distributions}
    \end{figure}

\begin{figure}[H]
        \centering
        \includegraphics[width=0.95\columnwidth]{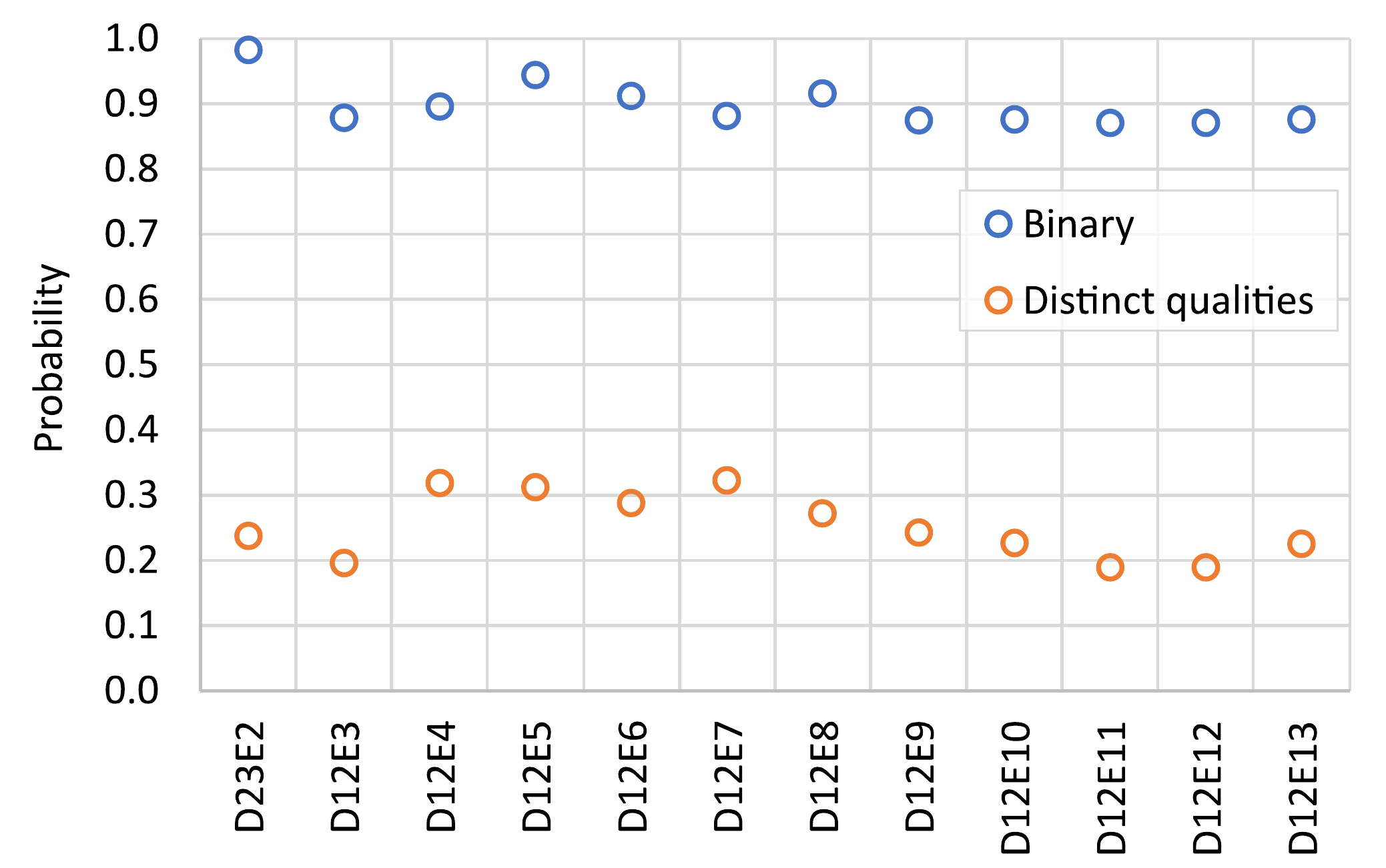}
        \caption{Amplified probabilities of a single marked vertex after repeated application of the same phase and walk parameters.}
        \label{fig:fixed_pairs_optimisation}
    \end{figure}

    In fact, there is another thing that is special about a binary marking function, which is not obvious from the results shown in \cref{fig:optimisation_landscape}. When applied to graphs in general (at least in the case of a single marked vertex), they appear to be capable of producing optimal amplification with repeated applications of the same phase and walk parameters. So in other words, a binary quality distribution is capable of effectively reducing arbitrarily large optimisation problems to only a 2 dimensional optimisation landscape. Refer to \cref{fig:fixed_pairs_binary} and \cref{fig:fixed_pairs_nonbinary} for a visual display of how the restricted 2D optimisation landscape produces high amplification in the binary case but is unable to in the case of a quality distribution without degeneracy (both for the 24-vertex complete graph and for 3 QWOA iterations). In addition, \cref{fig:fixed_pairs_optimisation} shows the optimised probabilities for the binary case contrasted against the no-degeneracy case (with uniformly distributed qualities) subject to $r=3$ repeated parameter pairs for a range of 24-vertex graphs and the result is consistent across all of them, not just the complete graph. Note that the optimisation procedure consists of producing 1,000 randomly generated sets of initial parameter pairs. The 10 parameter pairs that produce maximum initial amplification are taken as the initial values in a Nelder-Mead optimisation procedure. The most optimal of these is taken to be the final amplified probability for that particular graph and quality distribution. 
    
    All this is to say that the a binary marking function produces the highest amplification a single marked vertex for any graph, but also does so with repeated applications of the same parameter pairs that can be easily acquired via an optimisation procedure. The fact that repeated applications of the same parameter pairs can achieve maximum amplification also hints that it may be possible to find these parameters analytically. Out of all the graphs and quality distributions, the binary-marked complete graph seems to produce the highest amplifications. 
    
    \newpage
    
    \newpage
    
    \section{Algorithm pseudocode}
    \label{sec:pseudocode}
    
	The following algorithm finds a suitable quality threshold which produces a maximally amplified state for the final rotation count, $r_f$, where $r_f$ is a power of 2. Note that $\Psi_r(T)$ refers to the amplified state prepared via $r$ Grover rotations of a binary-marked solution space with quality threshold, $T$. Also note that the algorithm has been written for a minimisation problem.
	
	\begin{algorithmic}[1]
	    \Function{FinalThreshold}{$r_f$}    
            \State $r \gets 1$
            \State $sample \gets \text{200 randomly sampled solutions}$
            \State $median \gets \text{median of qualities in } sample$
            \State $Qu_1 \gets \text{Quartile 1 of qualities in } sample$
            \State $stepsize \gets (median-Qu_1)/10$
            \State $T \gets median$
            \State $T_1 \gets \textsc{FindPeak}(r,T,stepsize)$
            \State $r \gets 2$
            \State $T_2 \gets \textsc{FindPeak}(r,T_1,stepsize)$
            \State $r \gets 4$
            \While{$r<r_f$}
                \State $stepsize \gets (T_{r/4}-T_{r/2})/10$
                \State $T_r \gets \textsc{FindPeak}(r,T_{r/2},stepsize)$
                \State $r \gets 2*r$
            \EndWhile
            \State $stepsize \gets (T_{r/4}-T_{r/2})/10$
            \State $T \gets \textsc{ThresholdForAS}(r,T_{r/2},stepsize)$
            \State \Return $\textsc{AdaptiveSearch}(r,T)$
        \EndFunction
        \State
        \Function{FindPeak}{$r,T,stepsize$}
            \State $sum \gets 0$
            \State $weights \gets 0$
            \For{$i = 1\text{ to }20$}
                \State $T \gets T - stepsize$
                \State $count \gets 0$
                \While{$count<20$}
                    \State $x \gets \text{Measure}[\Psi_r(T)]$
                    \If{$f(x)<T$} \State $count \gets count + 1$
                    \Else
                        \State $sum \gets sum + T*count^4$
                        \State $weights \gets weights + count^4$
                        \State \textbf{break while loop}
                    \EndIf
                \EndWhile
                \If{$count = 20$} \State \Return T
                \EndIf
            \EndFor
            \State \Return $sum/weights$
        \EndFunction
        \State
        \Function{ThresholdForAS}{$r,T,stepsize$}
            \State $best \gets T$
            \For{$i = 1\text{ to }20$}
                \State $T \gets T - stepsize$
                \State $x \gets \text{Measure}[\Psi_r(T)]$
                \If{$f(x)<best$} \State $best \gets f(x)$
                \EndIf
            \EndFor
            \State \Return $best$
        \EndFunction
        \State
        \Function{AdaptiveSearch}{$r,T$}
            \State $count \gets 0$
            \While{$count < 40$}
                \State $count \gets 0$
                \State $hit \gets \text{False}$
                \While{$hit = \text{False}$}
                    \State $x \gets \text{Measure}[\Psi_r(T)]$
                    \State $count \gets count + 1$
                    \If{$f(x)<T$}
                        \State $hit = \text{True}$
                        \State $T \gets f(x)$
                    \EndIf
                \EndWhile
            \EndWhile
            \State \Return $T$
        \EndFunction
	\end{algorithmic}
    \hfill \break
    The output, $T$, of \textsc{FinalThreshold}$(r_f)$, is a threshold suitable for the second part of the MAOA, where the state, $\Psi_{r_f}(T)$, produces solutions in the marked set which are maximally amplified. As such, the user can repeatedly prepare and measure from this state in order to deliver maximum speedup with respect to finding optimal solutions within the marked set.
    
\end{document}